\documentclass{article}

\usepackage{orthodox}
\usepackage{makecell}

\setlist[itemize]{leftmargin=*}

\newtheorem{requirement}[theorem]{Requirement}

\renewcommand{\hat}{\widehat}

\setlist[itemize]{leftmargin=*}

\def\deg#1{\mathsf{d#1}}

\def \vcinfo#1#2{m_{#1\rightarrow #2}}
\def \cvinfo#1#2{\hat{m}_{#1\rightarrow #2}}
\def \vcinfoidx#1#2#3{m_{#1\rightarrow #2}^{(#3)}}
\def \cvinfoidx#1#2#3{\hat{m}_{#1\rightarrow #2}^{(#3)}}

\def \vcmean#1#2{\mu_{#1\rightarrow #2}}
\def \vcvar#1#2{v_{#1\rightarrow #2}}
\def \vcmeanidx#1#2#3{\mu_{#1\rightarrow #2}^{(#3)}}
\def \vcvaridx#1#2#3{v_{#1\rightarrow #2}^{(#3)}}

\def \cvmean#1#2{\hat{\mu}_{#1\rightarrow #2}}
\def \cvvar#1#2{\hat{v}_{#1\rightarrow #2}}
\def \cvmeanidx#1#2#3{\hat{\mu}_{#1\rightarrow #2}^{(#3)}}
\def \cvvaridx#1#2#3{\hat{v}_{#1\rightarrow #2}^{(#3)}}

\def \tilvcmeanidx#1#2#3{\wt{\mu}_{#1\rightarrow #2}^{(#3)}}
\def \tilvcvaridx#1#2#3{\wt{v}_{#1\rightarrow #2}^{(#3)}}

\title{\huge  \bfseries A General Compressive Sensing Construct using Density Evolution}

\author{ 
Hang Zhang, 
Afshin Abdi, 
and Faramarz Fekri
\\ 
\normalsize{School of Electrical and Computer Engineering, Georgia Institute of Technology, Atlanta, GA, USA.}
}

\date{}

\begin{document}
\maketitle

\begin{abstract}
This paper  \footnote{
Partial preliminary results appeared in 2021 IEEE Information Theory Workshop \cite{hzhang_density_evolve}.
} 
proposes a general framework to design 
a sparse sensing matrix 
$\ensuremath{\mathbf{A}}\in \mathbb{R}^{m\times n}$, 
for a linear measurement system
$\ensuremath{\mathbf{y}} = \ensuremath{\mathbf{Ax}}^{\natural} +
\ensuremath{\mathbf{w}}$, where $\ensuremath{\mathbf{y}} \in \mathbb{R}^m$,
$\ensuremath{\mathbf{x}}^{\natural}\in \RR^n$,
and $\ensuremath{\mathbf{w}}\in \RR^m$
denote the measurements, the signal with certain structures,
and the measurement noise, respectively.
By viewing the signal reconstruction from the measurements as a message passing algorithm 
over a graphical model,
we leverage tools from coding theory in the design of low density parity check codes,
namely the density evolution technique, and provide a framework for the design of matrix
$\ensuremath{\mathbf{A}}$. 
Two design schemes for the sensing matrix, namely, $(i)$ a regular sensing and $(ii)$ a
preferential sensing, are proposed and 
are incorporated into a single framework. 
As an illustration, we consider the $\ell_1$ regularizer,
which corresponds to Lasso for both of the schemes. 
In the regular sensing scenario, our framework can 
reproduce the classical result of Lasso, i.e., $m\geq c_0 k\log(n/k)$ after a proper distribution approximation, where $c_0 > 0$ is some fixed constant. 
In the preferential sensing scenario, we consider the 
case in which the whole signal is divided 
into two disjoint parts, namely, high-priority 
part $\bx^{\natural}_H$ and low-priority part $\bx^{\natural}_L$. 
Then, by formulating the sensing system design 
as a bi-convex optimization problem, we obtain 
a sensing matrix $\bA$ which provides a preferential treatment for 
$\bx^{\natural}_H$. Numerical 
experiments with both synthetic data and real-world data 
suggest a significant reduction of the 
$\ell_2$ error in the high-priority part $\bx^{\natural}_H$
and a slight reduction of the $\ell_2$ error in 
the whole signal $\bx^{\natural}$. 
Apart from the sparse signal with 
$\ell_1$ regularizer, our framework is flexible and 
capable of
adapting the design of the sensing matrix $\bA$ to a signal
$\ensuremath{\mathbf{x}}^{\natural}$ with other underlying structures. 
\end{abstract}

\section{Introduction}
This paper considers a linear sensing relation as  
\begin{align}
\label{eq:sense_relation}
\by = \bA\bx^{\natural} + \bw, 
\end{align}
where $\by \in \RR^m$ denotes the measurements, 
$\bA \in \RR^{m\times n}$ is the sensing matrix, 
$\bx^{\natural} \in \RR^n$ is the signal to be reconstructed, 
and $\bw\in \RR^m$ is the measurement noise 
with iid Gaussian distribution $\normdist(0, \sigma^2)$. 
To reconstruct $\bx^{\natural}$ from $\by$, one widely used method 
is the regularized M-estimator
\begin{align}
\label{eq:m_estim}
\wh{\bx} = \argmin_{\bx \in \RR^n}~\frac{1}{2\sigma^2}\norm{\by - \bA \bx}{2}^2 + f(\bx),
\end{align}
where $f(\cdot)$ is the regularizer used to enforce
a desired structure for $\wh{\bx}$. To ensure reliable recovery 
of $\bx^{\natural}$, sensing matrix $\bA$ needs to satisfy 
certain conditions, e.g., the incoherence in \cite{donoho2005stable}, 
RIP in \cite{candes2006stable, candes2006robust}, the \emph{neighborhood stability} in \cite{meinshausen2006high}, 
\emph{irrespresentable condition} in \cite{zhao2006model}, etc.
Notice that all the above works treat each entry of $\bx^{\natural}$ 
equally. However, 
in certain applications, entries of $\bx^{\natural}$ may have unequal 
importance from the recovery perspective. 
One practical application is the image compression, 
i.e.,  JPEG compression, where coefficients corresponding 
to the high-frequency part are more critical than
the rest of coefficients. \footnote{
An introduction can be found in \url{https://jpeg.org/jpeg/documentation.html}. 
}

In this work, we focus on the sparse sensing matrix $\bA$.
Leveraging tools from coding theory, namely, 
\emph{density evolution} (DE), we propose a heuristic but general design framework of $\bA$ 
to meet the requirements of the signal 
reconstruction such as placing more importance on the accuracy of a certain components of the signal. 
At the core of our work is the application of DE in 
\emph{message passing} (MP) algorithm, which is also referred to as belief propagation, or sum-product, or min-sum algorithm.
These different names are due to its broad spectrum of applications and its constant rediscovery in different fields. 
In physics, this 
algorithm existed no later than $1935$, when Bethe 
used a free-energy functional to approximate the partition function (cf. \cite{mezard2009information}). In the probabilistic inference, 
Pearl developed it in $1988$ for acyclic Bayesian networks and showed it 
leads to the exact inference \cite{pearl2014probabilistic}. 
The most interesting thing is its discovery in the coding theory.
In early 1960s, Gallager proposed 
sum-product algorithm 
to decode \emph{low density parity check} (LDPC) codes over graphs \cite{gallager1962low}. 
However, Gallagher work was almost forgotten and was rediscovered again 
in 90s \cite{berrou1996near, mceliece1998turbo}. 
Later \cite{richardson2001capacity} equipped it with DE
and used it for the design of LDPC codes
for capacity achieving over certain channels.

When narrowing down to the \emph{compressed sensing} (CS), MP has been widely used for
signal reconstruction  
\cite{sarvotham2006compressed, zhang2012verification,
kudekar2010effect, 
eftekhari2012density,
krzakala2012statistical, krzakala2012probabilistic, zdeborova2016statistical, 
donoho2009message, maleki2010approximate} 
and analyzing the performance under some specific sensing matrices.
The following briefly discusses the related work in 
the sensing matrix.

\par 
\noindent \textbf{Related work.}
In the context of the sparse sensing matrix,
the authors in \cite{sarvotham2006sudocodes}
first proposed a so-called sudocode construction technique and 
later presented a decoding algorithm based on
 the MP in \cite{baron2009bayesian}.
In \cite{chandar2010simple}, the non-negative sparse 
signal $\bx^{\natural}$ is considered under the 
binary sensing matrix. 
The work in \cite{dimakis2012ldpc} linked the channel encoding 
with the CS and 
presented a deterministic way of constructing 
sensing matrix based on a high-girth LDPC code. 
In \cite{luby2005verification, zhang2012verification, eftekhari2012density}, the authors considered the verification-based decoding and analyzed its performance 
with DE. 
In \cite{kudekar2010effect}, the spatial coupling is first introduced into CS and is evaluated 
with the decoding scheme adapted from 
\cite{luby2005verification}. 
However, all the above mentioned works focused 
on the noiseless setting, i.e., $\bw = \bZero$ in \eqref{eq:sense_relation}. 
In \cite{krzakala2012statistical, krzakala2012probabilistic, zdeborova2016statistical}, the noisy measurement 
is considered. A sparse sensing matrix based on 
spatial coupling is analyzed in the large system 
limit with replica method and DE. 
They proved its recovery performance to be optimal 
 when $m$ increases at the 
same rate of $n$, i.e., $m = O(n)$.

Moreover, in the context of a dense sensing matrix,  
the analytical tool switches from DE to 
\emph{state evolution} (SE), which is first proposed 
in \cite{donoho2009message, maleki2010approximate}. 
Together with SE comes the \emph{approximate message passing} (AMP)
decoding scheme. The empirical experiments suggest AMP has 
better scalability when compared with $\ell_1$ construction scheme without much scarifice in the performance. 
Additionally, 
an  exact phase transition formula can be obtained from SE, which 
predicts the performance of AMP to a good extent.
Later, \cite{bayati2011dynamics} provided
a rigorous proof for the phase transition 
property by the conditioning technique
from Erwin Bolthausen and \cite{donoho2016high} extended AMP to general M-estimation. 

Note that the above mentioned related works are not exhaustive due to 
their large volume. For a better understanding of the MP algorithm,   
the DE, and their application to the 
compressive sensing, 
we refer the interested 
readers to \cite{mezard2009information, montanari2012graphical, zdeborova2016statistical}. 
In addition to the work based on MP, there are other works based on 
 LDPC codes or graphical models. Since they 
 are not closely related to ours, we only 
 mention their names without further discussion
 \cite{xu2007efficient, xu2007further, khajehnejad2009nonnegative, jafarpour2009efficient, lu2012sparse, zhang2015deterministic, mousavi2017deepcodec}.

\par \noindent
\textbf{Contributions.}\
Compared to the previous works exploiting MP 
\cite{luby2005verification, zhang2012verification, eftekhari2012density, kudekar2010effect, krzakala2012statistical, krzakala2012probabilistic, zdeborova2016statistical, donoho2009message, maleki2010approximate}, our 
focus is on the sensing matrix design rather 
than the decoding scheme, which is based on the M-estimator with 
regularizer. 
Exploiting the DE, we 
propose a universal framework 
which supports both the 
regular sensing and the preferential sensing
for recovering the signal. 
Taking the sparse recovery with $\ell_1$ regularizer
as an example, we list our 
contributions as follows.

\begin{itemize}[noitemsep, topsep=0pt, leftmargin=*]
\item 
\textbf{Regular Sensing}. 
We consider the $k$-sparse signal $\bx^{\natural}\in \RR^n$
and associate it with a prior distribution 
such that each entry is zero with probability $1-k/n$. 
First we approximate this distribution 
with Laplacian prior by letting the probability mass near
the origin point to be $1-k/n$. 
Afterwards, 
we can reproduce the classical results in CS, i.e., 
$m\geq c_0 k\log n$.

\item 
\textbf{Preferential Sensing}. 
We design the sensing matrix that would result in 
more accurate (or exact) recovery of 
the high-priority sub-block of the signal 
relative to the low-priority sub-block. 
Numerical experiments suggest our framework can $(i)$   
reduce the error in the high-priority sub-block significantly; 
and $(ii)$ yet be able to reduce the error with regard to the whole signal modestly as well. 
Additionally, we emphasize that although we focus on two 
levels of priority in signal components in this 
work, we can easily extend the framework to 
the scenario where multiple levels of 
preferential treatment on the signal components are needed, 
by simply incorporating associated equations 
into the DE.
\end{itemize}

\noindent \textbf{Organization.}\ 
In Sec.~\ref{sec:prob_descrip}, we
formally state our problem and construct the graphical model.
In Sec.~\ref{sec:regular_design}, we focus on 
the regular sensing and propose the density 
evolution framework.
In Sec.~\ref{sec:unequal_protect}, the framework is further 
extended to the preferential sensing.  
Generalizations are presented 
in Sec.~\ref{sec:general}, 
simulation results are put in Sec.~\ref{sec:simul}, and 
conclusions are drawn in Sec.~\ref{sec:conclude}.

 \section{Problem Description}
\label{sec:prob_descrip}
We begin this section with a formal statement of our problem. 
Consider the linear measurement system
\vspace{-1mm} 
\begin{align*}
\by = \bA\bx^{\natural} + \bw, 
\vspace{-1mm} 
\end{align*}
where $\by\in \RR^m$, $\bA\in \RR^{m\times n}$, $\bx^{\natural}\in \RR^n$, 
and $\bw\in \RR^m$, respectively, denote the observations, 
the sensing matrix, the signal, 
and the additive sensing noise with 
its $i$th entry $w_i\iid \normdist(0, \sigma^2)$.
We would like to recover $\bx^{\natural}$ with the regularized 
M-estimator, which is written as 
\begin{align*}
\hat{\bx} = \argmin_{\bx} \frac{1}{2\sigma^2}\norm{\by - \bA\bx}{2}^2 + f(\bx), 
\end{align*}
where $f(\cdot)$ is the regularizer
used to enforce certain underlying structure 
for signal $\wh{\bx}$. 
Our goal is to 
design a sparse sensing matrix $\bA$ 
which provides preferential treatment 
for a sub-block of the signal $\bx^{\natural}$. 
In other words, 
the objective is to have a sub-block of the signal to
be recovered 
with lower probability of error when comparing with the 
rest of $\bx^{\natural}$. 
Before we proceed, we list our
two assumptions:

\begin{itemize}[leftmargin = *]
\item 
Measurement system $\bA$ is assumed to be sparse. 
Further, $\bA$ is assumed to have entries with
$\Expc A_{ij} = 0$, and $A_{ij} \in \{0, \pm A^{-1/2}\}$, 
where an entry $A_{ij}= A^{-1/2}$ (or $-A^{1/2}$) implies a positive (negative) relation between 
the $i$th sensor and the $j$th signal component. 
Having zero as entry implies no 
relation. 

\item 
The regularizer $f(\bx)$ is assumed to be separable such that 
$f(\bx) = \sum_{i=1}^n f_i(x_i)$. If it is 
not mentioned specifically, 
we assume all functions $f_i(\cdot)$ are the same. 

\end{itemize}

  First we transform
\eqref{eq:sense_relation} to a factor graph 
\cite{richardson2008modern}. 
Adopting the viewpoint of 
Bayesian reasoning, 
we can 
reinterpret M-estimator in \eqref{eq:m_estim}
as the \emph{maximum a posteriori} (MAP) estimator and rewrite it
as
\[
\hat{x} = \argmax_x \exp\bracket{-\frac{\norm{\by - \bA\bx}{2}^2}{2\sigma^2}}\times 
\exp\bracket{-f(\bx)}. 
\]
The first term $\exp\bracket{-\frac{\norm{\by - \bA\bx}{2}^2}{2\sigma^2}}$ is
viewed as the probability $\Prob(\by|\bx)$ while the second term
$\exp(-f(\bx))$ is regarded as the prior imposed on $\bx$.
Notice the term $e^{-f(\cdot)}$
may not necessarily be the true prior on
$\bx^{\natural}$.

As in \cite{montanari2012graphical}, we 
associate \eqref{eq:m_estim} with 
a factor graph $\calG = \bracket{\calV, \calE}$, where $\calV$ denotes the 
node set and $\calE$ is the edge set. 
First we discuss set $\calV$, which consists 
of two types of nodes: variable nodes and check nodes. 
We represent each entry $x_i$ as a variable node $v_i$ and each entry 
$y_a$ as a check node $c_a$. Additionally, we construct a check node 
corresponds to each  prior $e^{- f(x_i)}$. 
Then we construct the 
edge set $\calE$ by: $(i)$ placing 
an edge between the check node of the prior $e^{-f(x_i)}$ and the variable node
$v_i$, and $(ii)$ introducing an edge between the 
variable node 
 $v_i$ and $c_j$ iff 
$A_{ij}$ is non-zero. Thus, the design of $\bA$ 
is transformed to the problem of graph connectivity in $\calE$. 
Before to proceed, we list the notations used in this work.

\noindent \textbf{Notations.}\ 
We denote $c, c^{'}, c_0 > 0$ as some fixed positive constants. 
For two arbitrary real numbers $a, b$, 
we denote $a\lsim b$ when there exists some positive constant $c_0 > 0$ 
such that $a \leq c_0 b$. 
Similarly, we define the notation $a\gsim b$. 
If $a\lsim b$ and $a\gsim b$ hold simultaneously, we denote as 
$a\asymp b$. We have $a\propto b$ when $a$ is proportional to $b$. 
 For two distributions $d_1$ and $d_2$, we denote 
$d_1 \cong d_2$ if they are equal up to some normalization.

\section{Sensing Matrix for Regular Sensing}
\label{sec:regular_design}
\begin{figure}
\centering
\includegraphics[width = 3.4in]{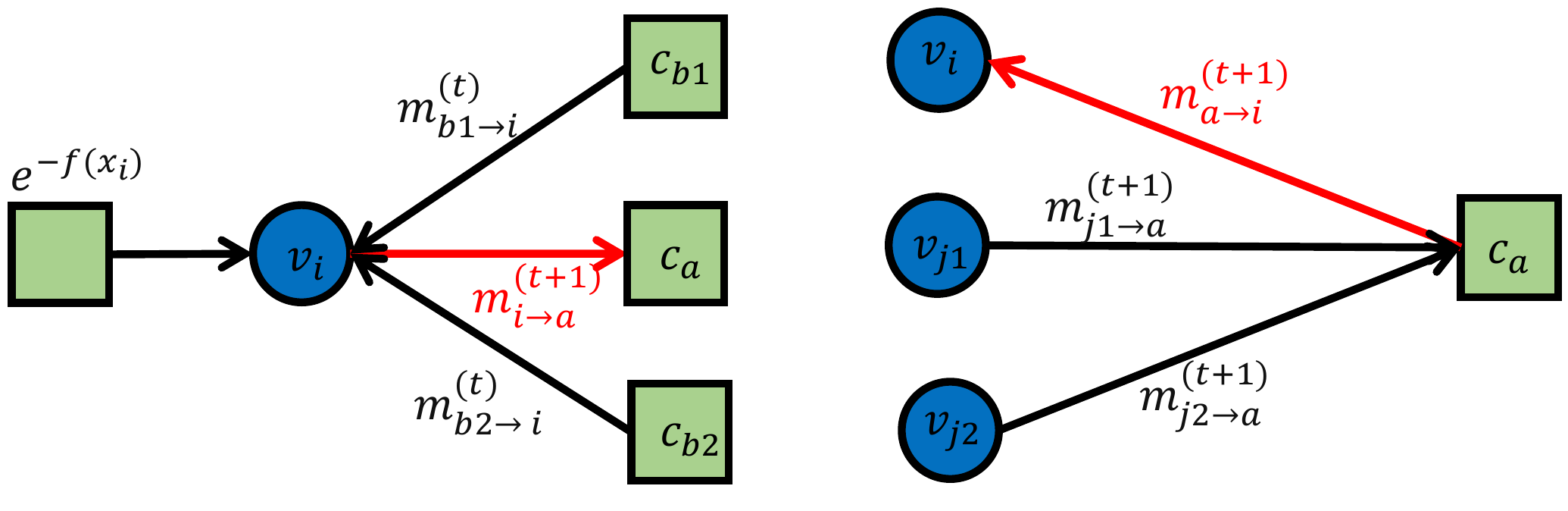}
\caption{Illustration of the message-passing algorithm, where 
the square icons represent the check nodes while the circle
icons represent the variable nodes.}
\label{fig:message_pass}	
\end{figure}

With the aforementioned graphical model, we can view recovering $\bx^{\natural}$
as an inference problem, which can be solved via the message-passing algorithm 
\cite{richardson2008modern}. 
Adopting the same 
notations as in \cite{montanari2012graphical}
as shown in Fig.~\ref{fig:message_pass}, we denote 
$\vcinfoidx{i}{a}{t}$ as the message from the variable node $v_i$ to check node $c_a$ 
at the $t^{\mathsf{th}}$ round of iteration.
Likewise, we denote $\cvinfoidx{a}{i}{t}$ as the 
message from the check node $c_a$ to variable node $v_i$.
Then message-passing algorithm is 
written as 
\begin{align} 
\label{eq:message_pass}
\vcinfoidx{i}{a}{t+1}(x_i) &\cong e^{-f(x_i)} \prod_{b\in \partial i \setminus a}
\cvinfoidx{b}{i}{t}(x_i); \\
\cvinfoidx{a}{i}{t+1}(x_i) &\cong \hspace{-1mm}\int\hspace{-2mm} \prod_{j\in \partial a\setminus i}  \vcinfoidx{j}{a}{t+1}(x_i) \cdot e^{-\frac{\bracket{y_a - \sum_{j=1}^n A_{aj}x_j}^2}{2\sigma^2}}dx_j,
\end{align}

\par \vspace{-2mm}
\noindent where $\partial i$ and $\partial a$ denote the neighbors connecting with 
$v_i$ and $c_a$, respectively, and 
the symbol $\cong$ refers to the equality up to the normalization. 
At the $t$th iteration, 
we recover $x_i$ by 
maximizing the posterior probability

{\small 
\vspace{-2mm}
\begin{align}
\label{eq:map_exp}
\hat{x}^{(t)}_i = \argmax_{x_i} \Prob(x_i|\by) \approx
\argmax_{x_i}e^{-f(x_i)} \prod_{a\in \partial i} \cvinfoidx{a}{i}{t}(x_i).
\end{align}
\vspace{-2mm}
}

\par
\begin{figure}
\centering
\includegraphics[width = 1in, height = 1in]{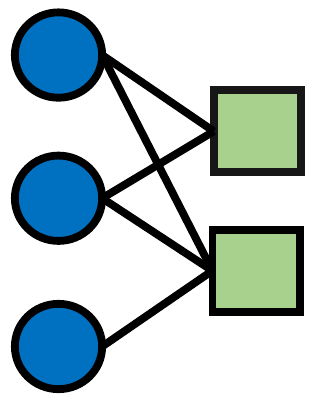}

\vspace{-2mm}
\caption{Illustration of the generating polynomials: 
$\lambda(\alpha) = \frac{1}{3}+ \frac{2\alpha}{3}$ and 
$\rho(\alpha)=\frac{\alpha}{2} + \frac{\alpha^2}{2}$. 
The square icons represent the check nodes while the 
circle icons represent the variable nodes.}
\label{fig:graph_distrib}	
\end{figure}

 \noindent
In the design of matrix $\bA$, 
there are some general desirable properties 
that we wish to hold (specific requirements will be discussed later): 
$(i)$ a correct signal reconstruction under the noiseless setting; 
and $(ii)$ minimum number of measurements, or equivalently minimum $m$. 
Before proceeding, we first introduce the generating polynomials 
$\lambda(\alpha) = \sum_{i}\lambda_i \alpha^{i-1}$ and 
$\rho(\alpha) = \sum_{i}\rho_i \alpha^{i-1}$, which correspond to the
degree distributions for  variable nodes and check nodes, 
respectively. We denote the coefficient $\lambda_i$ as the fraction of 
variable nodes with degree $i$, and similarly we define 
$\rho_i$ for the check nodes. 
An illustration of the generating polynomials 
$\lambda(\alpha)$ and $\rho(\alpha)$ is shown in 
Fig.~\ref{fig:graph_distrib}. 

\vspace{-3mm}

\subsection{Density evolution}
To design the matrix $\bA$, 
we study the reconstruction of $\bx^{\natural}$ 
from $\by$ via the convergence analysis 
of the message-passing over the factor graph. 
Due to the parsimonious 
setting of $\bA$, we have $\calE$ to be sparse and 
propose to borrow a tool known as \emph{density evolution} (DE)
\cite{richardson2001design, richardson2008modern, chung2000construction} 
that is already proven to be very powerful in  
analyzing the convergence in sparse graphs resulting from 
 LDPC.

Basically, DE views $\vcinfoidx{i}{a}{t}$ and $
\cvinfoidx{a}{i}{t}$
as RVs and tracks the changes of their probability distribution. 
When the message-passing algorithm converges, we would expect their 
distributions to become more concentrated.  
However, different from discrete RVs, 
continuous RVs $\vcinfoidx{i}{a}{t}$ and 
$\cvinfoidx{a}{i}{t}$ in our case require infinite bits for 
their precise representation in general, leading to complex formulas 
for DE.  
To handle such an issue, we approximate 
$\vcinfoidx{i}{a}{t}$ and 
$\cvinfoidx{a}{i}{t}$ as Gaussian RVs, i.e.,
 $\vcinfo{i}{a}\sim \normdist(\vcmean{i}{a}, \vcvar{i}{a})$ and $\cvinfo{a}{i}\sim\normdist(\cvmean{a}{i}, \cvvar{a}{i})$, respectively.
Since the Gaussian distribution is uniquely determined by its mean 
and variance, we will be able to reduce 
the DE to finite dimensions as in 
\cite{chung2000construction, krzakala2012statistical, krzakala2012probabilistic}.

In our work, the DE tracks two quantities 
$E^{(t)}$ and $V^{(t)}$,
which denote the deviation from the 
mean and average of the variance, respectively, 
and are defined as

{\small 
\vspace{-4mm}
\[
E^{(t)} &= \frac{1}{m\cdot n}\sum_{i=1}^n \sum_{a=1}^m \bracket{\vcmean{i}{a}^{(t)} - x^{\natural}_i}^2; \\
V^{(t)} &= \frac{1}{m\cdot n}\sum_{i=1}^n \sum_{a=1}^m \vcvar{i}{a}^{(t)}.
\]
}

\par \vspace{-2mm}\noindent
Then we can show that the DE analysis 
yields

{\small \vspace{-2mm}
\begin{align}
\label{eq:de_equal_protect}
E^{(t+1)} = \Expc_{\prior(s)}\Expc_{z} \bigg[h_{\mean}\bigg(& s + 
\sum_{i, j} \rho_{i} \lambda_{j}
z\sqrt{\frac{i}{j}E^{(t)} + \frac{A\sigma^2}{j}}; \notag \\
& \sum_{i,j}\rho_{i}\lambda_{j}
\frac{A\sigma^2 + i V^{(t)}}{j}\bigg)- s ]^2 ;  \\
V^{(t+1)} = \Expc_{\prior(s)}\Expc_z h_{\var}\bigg(& s + 
\sum_{i, j} \rho_{i} \lambda_{j}z
\sqrt{\frac{i}{j}E^{(t)} + \frac{A\sigma^2}{j}}; \notag\\ 
& \sum_{i,j}\rho_{i}\lambda_{j}
\frac{A\sigma^2 + i V^{(t)}}{j}\bigg),
\end{align}
}

\par \vspace{-2mm} \noindent
where $\prior(\cdot)$ denotes the true prior on
the entries of $\bx^{\natural}$, and $z$ is a standard normal RV $\normdist(0, 1)$.
The functions $h_{\mean}(\cdot)$ and $h_{\var}(\cdot)$ are to approximate
the mean $\vcmean{i}{a}$ and variance $\vcvar{i}{a}$, which are given by 

\vspace{-4mm}
{\small 
\begin{align}
\label{eq:hmeanvar_def}
  &h_{\mean}(\mu; v) \hspace{-1mm}=\hspace{-1mm} \lim_{\gamma \rightarrow \infty} \frac{\int x_i e^{-\gamma f(x_i)}
    e^{-\frac{\gamma\bracket{x_i-\mu}^2}{2v} } dx_i}{
    \int e^{-\gamma f(x_i)}
    e^{-\frac{\gamma\bracket{x_i-\mu}^2}{2v} } dx_i}; \\
&h_{\var}(\mu; v) \hspace{-1mm}= \hspace{-1mm} \lim_{\gamma \rightarrow \infty}\frac{\gamma\int x^2_i e^{-\gamma f(x_i)}
    e^{-\frac{\gamma\bracket{x_i-\mu}^2}{2v} } dx_i}{\int e^{-\gamma f(x_i)}
    e^{-\frac{\gamma\bracket{x_i-\mu}^2}{2v} } dx_i}- \bracket{h_{\mean}(\mu; v)}^2.  \notag 
\end{align}
}

\par \vspace{-4mm} \noindent
For detailed explanations and the proof, we refer interested readers to the supplementary material.


%

\subsection{Sensing matrix design} 
Once the values of polynomial coefficients
 $\set{\lambda_i}_i$ and 
$\set{\rho_i}_i$ are determined,  
we can construct a random graph $\calG = (\calV, \calE)$, or equivalently the
sensing matrix $\bA$, by setting $A_{ij}$ as 
$\Prob(A_{ij} = A^{-1/2}) = \Prob(A_{ij} = -A^{-1/2}) = \frac{1}{2}$,  
if there is an edge $(v_i, c_j) \in \calE$; otherwise we set $A_{ij}$
to zero. Hence the sensing matrix construction reduces 
to obtaining the feasible values of $\set{\lambda_i}_i$ and 
$\set{\rho_i}_i$ while satisfying 
certain properties for the signal reconstruction
as discussed in the following.

Our first requirement is to have a perfect signal reconstruction under the noiseless 
scenario ($\sigma^2 = 0$). This implies that 
\begin{itemize}
\item 
the algorithm must converge, i.e., 
$\lim_{t\rightarrow \infty} V^{(t)} = 0$;
\item 
the average error should shrink to zero, i.e., $\lim_{t\rightarrow \infty} E^{(t)} = 0$.
\end{itemize}
Second, we wish to minimize the number of measurements. Using the 
fact that $n\bracket{\sum_i i \lambda_i} = m\bracket{\sum_i i \rho_i} = \sum_{i, j}\Ind((v_i, c_j)\in \calE)$, we 
formulate the above two design criteria as the following optimization problem

\vspace{-2mm}
{\small 
\begin{align}
\label{eqn:density_evolve_optim}
\min_{\substack{\blambda \in \Delta_{\deg{v}_{\mathsf{max}} - 1}; \\\brho \in \Delta_{\deg{c}_{\mathsf{max}} - 1}}}~\
& \dfrac{m}{n} = \dfrac{\sum_{i\geq 2} i\lambda_{i}}{\sum_{i\geq 2} i \rho_{i}}, \\
\St~&~\lim_{t\rightarrow \infty}(E^{(t)}, V^{(t)}) = (0, 0); 
\label {eq:density_evolve_regular_converge_requirement}\\
 &~\lambda_1 = \rho_1 = 0, \label {eq:density_evolve_regular_converge_oneway} 
\end{align}
}

\par\vspace{-6mm}\noindent
where $\Delta_{d-1}$ denotes the $d$-dimensional simplex, namely, 
$\Delta_{d-1} \defequal \{\bz\in \RR^d~|~ \sum_i z_i = 1, z_i \geq 0\}$.
The constraint in 
\eqref{eq:density_evolve_regular_converge_oneway} is to avoid one-way message
passing as in \cite{chung2000construction, richardson2001capacity}.

Generally speaking, we need to run DE numerically 
to check the requirement \eqref{eq:density_evolve_regular_converge_requirement} 
for every possible values of 
$\set{\lambda_i}_i$ and $\set{\rho_i}_i$. 
However, for certain choices of regularizers $f(\cdot)$, we
can reduce the requirement \eqref{eq:density_evolve_regular_converge_requirement}
to a closed-form equation. 
As an example, we set the prior in \eqref{eq:message_pass} to be
a Laplacian distribution, i.e., $e^{-\beta|x|}$. 
In this case, the regularizer $f(\cdot)$ in \eqref{eq:m_estim}
becomes $\beta \norm{\cdot}{1}$ and the M-estimator  in 
\eqref{eq:m_estim} transforms to Lasso \cite{tibshirani1996regression}.

\vspace{-3mm}
\subsection{Example of regular sensing with a Laplacian prior} 
\label{subsec:equal_protect_laplacian}
Assuming the signal $\bx^{\natural}$ is $k$-sparse, i.e., 
$\norm{\bx^{\natural}}{0}\leq k$, we would like to recover 
$\bx^{\natural}$ with the regularizers $\beta\norm{\cdot}{1}$.
Following the approaches in \cite{donoho2009message} in the noiseless case, we can show that

\vspace{-4mm}
{\small
\begin{align}
\label{eq:de_lasso}
E^{(t+1)} =~& \Expc_{\prior(s)}\Expc_{z\sim \normdist(0, 1)}
\Bracket{\prox\bracket{s + a_1z\sqrt{E^{(t)}} ; \beta a_2 V^{(t)}}- s}^2; \notag \\
V^{(t+1)} =~&  \Expc_{\prior(s)}\Expc_{z\sim \normdist(0, 1)} \notag \\
\cdot &
\Bracket{ \beta a_2 V^{(t)} \prox^{'}\bracket{s + a_1z\sqrt{E^{(t)}}; \beta a_2 V^{(t)} }}, 
\end{align}
}

\vspace{-4mm}\par \noindent
where 
$a_1$ is defined as $\sum_{i, j} \rho_{i} \lambda_{j}
\sqrt{i/j}$, and $a_2$ is defined 
as $\sum_{i,j}\rho_{i}\lambda_{j} \bracket{i/j}$. Further, operator $\prox(a; b)
$ is the soft-thresholding estimator defined as $\sign(a)\max(|a| - b, 0)$, 
and operator $\prox^{'}(a; b)$ is the derivative w.r.t. the first argument.

\begin{remark} 
{
Unlike SE that only tracks $E^{(t)}$ \cite{donoho2009message}, our DE 
takes into account both the average variance $V^{(t)}$ and the deviation 
from mean $E^{(t)}$. 
Assuming $V^{(t)} \propto \sqrt{E^{(t)}}$, 
our DE equation w.r.t. $E^{(t)}$ in \eqref{eq:de_lasso}
reduces to a similar form as SE.
} 
\end{remark}

{
Having discussed its relation with SE, we now show 
that our DE can reproduce the classical results in 
compressive sensing, namely, $m\geq c_0 k\log(n/k) = O(k\log n)$
(cf. \cite{foucart2011hard})
under the regular sensing matrix design, i.e., when all 
variable nodes have the same degree $\deg{v}$ and 
the check nodes have the same degree $\deg{c}$. 
Before we proceed, we first approximate the 
ground-truth distribution with the Laplacian prior. 
Assuming that the entries of $\bx^{\natural}$ are iid
and $\bx^{\natural}\in \RR^n$ is $k$-sparse,
each entry becomes zero with probability $\bracket{1-k/n}$. 
Hence we set $\beta$ such that 
the probability mass within the region $[-c_0, c_0]$ 
(where $c_0$ is some small positive constant)
with the Laplacian prior is equal to $1-k/n$. That is  
}

{
\vspace{-3mm}
{\small 
\[
\frac{\beta}{2}\int_{|\alpha | \leq c_0}e^{-\beta |\alpha|}d\alpha = 1 - \frac{k}{n}. 
\]
}}

\par \vspace{-2mm} \noindent
This results in $\beta = n/(c_0 n \log(n/k))$. 
Then we conclude the following

\begin{theorem}
\label{thm:equal_lasso}
{
Let $\bx^{\natural}$ be a $k$-sparse signal
and assume that $\beta$ is set to $n/(c_0\log\bracket{n/k})$. 
Then, the necessary conditions for 
$\lim_{t\rightarrow \infty}\bracket{E^{(t)}, V^{(t)}} = 
\bracket{0, 0}$ in \eqref{eq:de_lasso} results in
$a_1^2\leq n/k$ and $a_2 \leq n/\bracket{c_0 k\log(n/k)}$, 
where $a_1$ and $a_2$ are defined as 
 $\sum_{i, j} \rho_{i} \lambda_{j}
\sqrt{i/j}$ and $\sum_{i,j}\rho_{i}\lambda_{j} \bracket{i/j}$, 
respectively. 
}
\end{theorem}

When turning to the regular design, namely,  
all variable nodes are with the degree $\deg{v}$
and likewise all check nodes are with degree $\deg{c}$, we can write  
$a_1$ and $a_2$ as  $\sqrt{n/m}$ and $n/m$, respectively. 
Invoking Thm.~\ref{thm:equal_lasso} will then yield the 
classical result of the
lower bound on the number of measurements  $m\geq c_0 k\log(n/k)$. 
The technical details are deferred to 
Sec.~\ref{thm_proof:equal_lasso}. 
In addition to the Laplacian prior, we also considered the 
Gaussian prior, i.e., $e^{-\beta \norm{\bx}{2}^2}$, which 
makes the M-estimator in \eqref{eq:m_estim} 
the ridge regression \cite{hastie01statisticallearning}. Corresponding discussion is left to 
Sec.~\ref{sec:gaussian_prior} for interested readers.

\section{Sensing Matrix for Preferential Sensing}
\label{sec:unequal_protect}
Having discussed the regular sensing scheme, 
%
%
this section explains as to how we apply our DE framework to design the sensing matrix 
$\bA$ such that we can provide preferential
treatment for different entries of $\bx^{\natural}$.
For example, the high priority components will be recovered more accurately than the low priority parts of $\bx^{\natural}$.

\subsection{Density evolution}
Dividing the entire $\bx^{\natural}$ into the 
high-priority part $\bx^{\natural}_{H} 
\in \RR^{n_H}$ and 
low-priority part 
$\bx^{\natural}_{L} \in \RR^{n_L}$,  
we separately introduce the generating polynomials 
$\lambda_{H}(\alpha) = \sum \lambda_{H, i}\alpha^{i-1}$
and 
$\lambda_L(\alpha) = \sum \lambda_{L, i} \alpha^{i-1}$
for the high-priority part
 $\bx^{\natural}_H$ and the 
 low-priority part $\bx^{\natural}_L$, 
respectively. Note that
$\lambda_{H, i}$ (and likewise $\lambda_{L, i}$)
denotes the fraction of variable nodes corresponding to 
high-priority part (low-priority part) with degree $i$. 
Similarly, we introduce the 
generating polynomials 
$\rho_{H}(\alpha) = \sum_i\rho_{H, i} \alpha^{i-1}$ 
and $\rho_{L}(\alpha) = \sum_i\rho_{L, i} \alpha^{i-1}$ 
for the edges of the check nodes connecting to the
high-priority part $\bx^{\natural}_H$ and to
the low-priority part $\bx^{\natural}_L$, 
respectively. 

Generalizing the analysis of the regular sensing, we
separately track the average error and variance for $
\bx^{\natural}_H$ and $\bx^{\natural}_L$. 
For the high-priority part $\bx^{\natural}_H$, 
we define $E_H$ as $\sum_{m}\sum_{i\in H} (\vcmean{i}{a}-x^{\natural}_i)^2/(m \cdot n_H)$
and $V_H$ as $\sum_{m}\sum_{i\in H} \vcvar{i}{a}/(m \cdot n_H)$, 
where $n_H$ denotes the length of the high-priority 
part $\bx^{\natural}_H$.
Analogously we define $E_L$ and $V_L$ for the 
low-priority part $\bx^{\natural}_L$. 
We then write the corresponding DE as

{\small \vspace{-4mm}
\begin{align}
\label{eq:unequal_de}
E_H^{(t+1)} &=\Expc_{\prior(s)}\Expc_{z\sim \normdist(0,1)}
\bigg[h_{\mean}\bigg( s + z \cdot b^{(t)}_{H, 1}; b^{(t)}_{H, 2}\bigg)- s\bigg]^2; \notag \\
V^{(t+1)}_{H} &=  \Expc_{\prior(s)}
\Expc_{z\sim \normdist(0, 1)} \bigg[h_{\var}\bigg(  s +z \cdot b^{(t)}_{H, 1}; b^{(t)}_{H,2}\bigg)\bigg], 
\end{align}
\vspace{-2mm}
}

\par \noindent
where $b_{H, 1}^{(t)}$
and $b_{H, 2}^{(t)}$ are defined as 

{\small \vspace{-2mm}
\begin{align*}
b^{(t)}_{H,1} &= 
\sum_{\ell, i, j}
\lambda_{H, \ell} \rho_{L, i} \rho_{H, j}
\sqrt{\frac{A\sigma^2+i E^{(t)}_L + j E^{(t)}_H}{\ell}}; \\
b^{(t)}_{H,2} &= 
\sum_{\ell, i, j}
\lambda_{H, \ell} \rho_{L, i} \rho_{H, j} \frac{A\sigma^2 + i V^{(t)}_L + j V^{(t)}_H}{\ell}.
\end{align*}
\vspace{-2mm}
}

\par \noindent
The definitions of 
$h_{\mean}$ and $h_{\var}$ are as in 
\eqref{eq:hmeanvar_def}.
Switching the index $H$ with $L$ yields the 
DE w.r.t. the pair 
$(E_L^{(t+1)}, V_L^{(t+1)})$.
Notice we can also put different regularizers 
$f_H(\cdot)$ and
$f_L(\cdot)$ for 
$\bx^{\natural}_H$ and $\bx^{\natural}_L$.
In this case, we need to modify the 
regularizers $f(\cdot)$ in \eqref{eq:hmeanvar_def} 
to $f_H(\cdot)$ and $f_L(\cdot)$, respectively. 

\subsection{Sensing matrix design}
In addition to the constraints used in  
\eqref{eqn:density_evolve_optim}, the 
sensing matrix for preferential sensing 
must satisfy the following constraint:
\par 
\noindent
\textbf{Consistency requirement w.r.t. edge number}. 
Consider the total number of edges incident with the high-priority part 
$\bx^{\natural}_H$,  
$\sum_{i\in H}\Ind\bracket{(v_i, c_a)\in \calE}$.
From the viewpoint of the variable nodes, we can 
compute this number as $n_H\bracket{\sum_i i\lambda_{H, i}}$.
Likewise, from the viewpoint of the check nodes, 
the total number of edges is obtained as 
$\sum_{i\in H}\Ind\bracket{(v_i, c_a)\in \calE} = m\bracket{\sum_i i\rho_{H, i}}$. 
Since the edge number should be the same with either
of the above two counting methods, we obtain 
\[
\sum_{i\in H}\Ind\Bracket{(v_i, c_a)\in \calE} = 
n_H\bracket{\sum_i i\lambda_{H, i}} 
= m\bracket{\sum_i i\rho_{H, i}}. 
\]
Similarly, the consistency requirement for the 
edges connecting to the low-priority part $\bx^{\natural}_L$
would give 
$\sum_{i\in L}\Ind\bracket{(v_i, c_a)\in \calE} = m (\sum_i i\rho_{L, i}) = 
n_L(\sum_{i}i \lambda_{L, i})$. 
\par 
Moreover, we may have additional constraints depending 
on the measurement noise: 
\begin{itemize}
\item 
\textbf{Preferential sensing for the noiseless measurement.}
In the noiseless setting ($\sigma = 0$), we require
$V_H$ and $V_L$ to diminish to zero to ensure 
the convergence of the MP algorithm. 
Besides, we require the 
average error $E_{H}^{(t)}$ in
the  high-priority part 
$\bx_H^{\natural}$ to be zero. Therefore, the requirements can be  
summarized as 
\begin{requirement}
\label{require:noiseless_unequal}
In the noiseless setting, i.e., $\sigma = 0$,
we require the quantities 
$E_H^{(t)}, V_H^{(t)}$, and $V_L^{(t)}$ in 
\eqref{eq:unequal_de} converge to zero
\begin{align}
\label{eq:unequal_de_require}
\lim_{t\rightarrow \infty}\big(E_H^{(t)}, V_H^{(t)}, V_L^{(t)}\big)= 
\bracket{0, 0, 0},
\end{align} 
which implies the MP converges and 
the high-priority part $\bx^{\natural}_H$
can be perfectly reconstructed. 
\end{requirement}
Notice that no constraint is placed on  
the average error $E_{L}^{(t)}$ for the low-priority 
part $\bx^{\natural}_L$, since it is given a lower priority 
in reconstruction. 
\item 

\textbf{Preferential sensing for the noisy measurement.}
Different from the noiseless setting, the high-priority 
part $\bx^{\natural}_H$ cannot be perfectly reconstructed in the presence of 
measurement noise, i.e., $\lim_{t\rightarrow \infty}E^{(t)}_H > 0$. Instead we consider the 
difference across iterations, namely, 
$\delta^{(t)}_{E, H} = E^{(t+1)}_{H} - E^{(t)}_{H}$ and 
$ \delta^{(t)}_{E, L} = E^{(t+1)}_{L} - E^{(t)}_{L}$, 
which corresponds to the convergence rate. 
To provide an extra protection for the high-priority part $\bx^{\natural}_H$, we would like 
$\delta^{(t)}_{H}$ to decrease at a faster rate. Hence, 
the following requirement:
\begin{requirement}
\label{require:noisy_unequal}
There exits a positive constant $T_0$ 
such that the average error $E^{(t)}_H$
converges faster than $E^{(t)}_L$ whenever $t\geq T_0$, 
i.e., $\abs{\delta^{(t)}_{E, H}} \leq \abs{ \delta^{(t)}_{E, L}}$.   
\end{requirement} 
\end{itemize}
Apart from the above constraints, we also require $\lambda_{L, 1} = \lambda_{H, 1} = \rho_{L, 1} = \rho_{H, 1}= 0$ to avoid one-way message passing \cite{richardson2001capacity, richardson2008modern, chung2000construction}.
Summarizing the above discussion, the 
design of the sensing matrix $\bA$
for minimum number of measurements
$m$ reduces to the following optimization problem  
\begin{align}
\label{eq:unequal_protect_optim_obj}
\min_{\substack{\blambda_L \in \Delta_{\deg{v}_{\mathsf{L}} - 1},\\ \blambda_H \in \Delta_{\deg{v}_{\mathsf{H}} - 1}, \\
 \brho_L \in \Delta_{\deg{c}_{\mathsf{H}} - 1}, \\
 \brho_H \in \Delta_{\deg{c}_{\mathsf{L}} - 1} }}&
\frac{m}{n} = \frac{{n_L}\bracket{\sum_i i \lambda_{L,i}} +
{n_H} \bracket{\sum_i i\lambda_{H, i}}}{
\sum_{i} i\bracket{\rho_{L, i} +\rho_{H, i}}}; \\
\St~~& \frac{\sum_i i \lambda_{L, i}}{\sum_i i \lambda_{H, i}} \times \
\frac{\sum_i i \rho_{H, i}}{\sum_i i \rho_{L, i}}  = \frac{n_H}{n_L}; 
\label{eq:unequal_degree_same}\\
& \textup{Requirement }\eqref{require:noiseless_unequal}
\textup{ and }\eqref{require:noisy_unequal};  \label{eq:unequal_protect_require} \\
& \lambda_{L, 1} = \lambda_{H, 1} = \rho_{ L, 1}= \rho_{H, 1}= 0,  
\end{align}
where $\Delta_{d-1}$ 
denotes the $d$-dimensional simplex,  and the parameters
$\deg{v}_{\mathsf{H}}$ and $\deg{c}_{\mathsf{L}}$ denote the 
maximum degree w.r.t.  the variable nodes corresponding 
to the high-priority part $\bx^{\natural}_H$ and 
low-priority part $\bx^{\natural}_{L}$, respectively. 
Similarly we define the maximum degree $\deg{c}_{\mathsf{H}}$ and $\deg{c}_{\mathsf{L}}$
w.r.t the check nodes.

The difficulties of the optimization 
problem in \eqref{eq:unequal_protect_optim_obj} 
come from two-fold: 
$(i)$ requirements
 from DE; 
 and $(ii)$ non-convex nature of 
\eqref{eq:unequal_protect_optim_obj}.
In the following scenario, we will revisit the example of
$\ell_1$ regularizer and show how to simplify the 
optimization problem in \eqref{eq:unequal_protect_optim_obj}.

\vspace{-2mm}
\subsection{Example of preferential sensing with a Laplacian prior}
\label{subsec:preferential_laplace_prior}

Consider a sparse signal $\bx^{\natural}$
whose high-priority part $\bx^{\natural}_H\in \RR^{n_H}$
and the low-priority part $\bx^{\natural}_L \in \RR^{n_L}$ 
are $k_H$-sparse and $k_L$-sparse, respectively. In addition, we assume $\frac{k_H}{n_H}\gg \frac{k_L}{n_L}$, 
implying that the high-priority part $\bx^{\natural}_H$ 
contains more data. 

Ideally, 
we need to numerically run the DE update 
equation in \eqref{eq:unequal_de} to 
check whether the requirement in 
\eqref{eq:unequal_protect_require} holds or not, which can be 
computationally prohibitive. 
In practice, we would relax these conditions 
to arrive at some closed forms. The following 
 outlines 
our relaxation strategy with all 
technical details being deferred to the supplementary material. 

\par \noindent
\textbf{Relaxation of Requirement \ref{require:noiseless_unequal}.}
First we require the variance to converge to zero, 
i.e., $\lim_{t\rightarrow \infty}(V^{(t)}_H, V_L^{(t)}) = 
\bracket{0, 0}$. The derivation of its necessary condition
consists of two parts: 
$(i)$ we require the point $\bracket{0, 0}$ to be a fixed point 
of the DE equation w.r.t. $V^{(t)}_H$ and 
$V^{(t)}_L$; and $(ii)$ we require that the average variance 
$(V^{(t)}_H, V^{(t)}_L)$ to converge in the region 
where the magnitudes of $V^{(t)}_H$ and
$V^{(t)}_L$ are sufficiently small.

The main technical challenge lies in investigating the
convergence of $(V^{(t)}_H, V^{(t)}_L)$. 
Define the difference $\delta_{V, H}^{(t)}$
and $\delta_{V, L}^{(t)}$
across iterations as $\delta_{V, H}^{(t)} \defequal V^{(t+1)}_H - V^{(t)}_H$ and  $\delta_{V, L}^{(t)} \defequal V^{(t+1)}_L - V^{(t)}_L$, respectively. Then, we  
obtain a linear equation 
\vspace{-2mm}
\begin{align*}
\begin{bmatrix}
\delta_V^{(H)}(t+1) \\
\delta_V^{(L)}(t+1)	
\end{bmatrix}  = 
\bL_V^{(t)}\begin{bmatrix}
\delta_V^{(H)}(t) \\
\delta_V^{(L)}(t)		
\end{bmatrix}
\end{align*}
via the Taylor-expansion. 
Imposing the convergence constraints on
$V_H^{(t)}$ and $V_L^{(t)}$, i.e., 
$\lim_{t\rightarrow \infty}\bracket{\delta^{(t)}_{V, H},\delta^{(t)}_{V, L} } = \bracket{0, 0}$, 
yields the condition $\inf_t \opnorm{\bL^{(t)}_{V}} \leq 1$.
That is 

{\small \vspace{-2mm}
\begin{align}
\label{eq:unequal_variance}
& \Bracket{
\bracket{\frac{\beta_H k_H}{n_H} 
\sum_{\ell}\frac{\lambda_{H, \ell}}{\ell}}^2 + 
\bracket{
\frac{\beta_L k_L}{n_L}
\sum_{\ell}\frac{\lambda_{L, \ell}}{\ell}}^2} \notag \\
\times ~& 
\Bracket{\bracket{\sum_i i\rho_{H, i}}^2 + \bracket{\sum_i i\rho_{L, i}}^2} 
\leq 1. 
\end{align} 
\vspace{-2mm}
}

\par \noindent
Then we turn to the behavior of $E^{(t)}_H$. Assuming $E^{(t)}_L$ converges to a fixed non-negative
constant $E_L^{(\infty)}$, we would like 
$E^{(t)}_H$ to converge to zero.  
Following the same strategy as above, 
we obtain the following condition 

{\small \vspace{-3mm}
\begin{align}
\label{eq:unequal_error}
\hspace{-3mm}\frac{k_H}{n_H}\bracket{\sum_{\ell}\frac{\lambda_{H, \ell}}{\sqrt{\ell}}}^2
\bigg[\bigg(\sum_i \sqrt{i}\rho_{H, i}\bigg)^2 
\hspace{-1mm}+\hspace{-1mm} \
\bigg(\sum_i \sqrt{i}\rho_{L, i}\bigg)^2
\bigg] \leq 1.
\end{align}
\vspace{-2mm}
}

\par \noindent
The technical details are
put in the supplementary material.
\par \noindent
\textbf{Relaxation of Requirement \ref{require:noisy_unequal}.}
First we define the difference across iterations 
as $\delta_{E, H}^{(t)} = E^{(t+1)}_H - E^{(t)}_H$ and 
$  \delta^{(t)}_{E, L} = E^{(t+1)}_L - E^{(t)}_L$.
Using the Requirement  \ref{require:noisy_unequal}, 
we perform the Taylor expansion w.r.t. the 
difference $\delta_{E, H}^{(t)}$ and
$\delta^{(t)}_{E, L}$, 
and obtain the linear equation 
\[
\begin{bmatrix}
\delta_{E, H}^{(t+1)} \\
\delta^{(t+1)}_{E, L}
\end{bmatrix} = 
\begin{bmatrix}
L_{E, 11} & L_{E, 12} \\
L_{E, 21} & L_{E, 22}
\end{bmatrix}
\begin{bmatrix}
\delta_{E, H}^{(t)} \\
 \delta^{(t)}_{E, L}
\end{bmatrix} .
\]
To ensure the reduction of $\delta_{E, H}^{(t)}$ 
at a faster rate than $\delta_{E, L}^{(t)}$, 
we would require 
$L_{E, 11} \leq L_{E, 21}$
and 
$L_{E, 12} \leq L_{E, 22}$.
This results in

{\small \vspace{-4mm}
\begin{align}
\label{eq:unequal_error_decrease_faster}
\frac{k_H}{n_H}\bracket{\sum_{\ell}\frac{\lambda_{H, \ell}}{\sqrt{\ell}}}^2 
\leq 
\frac{k_L}{n_L}\bracket{\sum_{\ell}\frac{\lambda_{L, \ell}}{\sqrt{\ell}}}^2. 
\end{align}
\vspace{-2mm}
}

\par \noindent
Summarizing the above discussion, we transform the constraint 
in 
\eqref{eq:unequal_protect_require} to  the closed-form and 
find the local optimum of \eqref{eq:unequal_protect_optim_obj}
via an alternating minimization method.

\section{Potential Generalizations}
\label{sec:general}
This section discusses two possible 
generalizations, i.e., 
non-exponential family priors and reconstruction via 
a \emph{minimum mean square error} (MMSE) decoder.
The design principles of the sensing matrix are exactly the same
as \eqref{eqn:density_evolve_optim} and
 \eqref{eq:unequal_protect_optim_obj}
except that the DE equations need to be modified. 

\subsection{Non-exponential priors}
Previous sections assume the prior 
to be $e^{-f(\bx)}$, which belongs to the
exponential family distributions. In 
this subsection, we  generalize it to 
arbitrary distributions $\wh{\prior}(\bx)$. 
One example of the non-exponential distribution is   
sparse Gaussian, i.e., $\frac{k}{n} e^{-(x - \mu)^2/2\sigma^2} + \bracket{1-\frac{k}{n}}
\delta(x)$, which is used to model $k$-sparse signals.
With the generalized prior, the MP in \eqref{eq:message_pass} 
is modified to 
\begin{align}
\label{eq:general_message_pass}
\vcinfoidx{i}{a}{t+1}(x_i) &\cong \wh{\prior}(x_i) \prod_{b\in \partial i \setminus a}
\cvinfoidx{b}{i}{t}(x_i); \notag \\
\cvinfoidx{a}{i}{t+1}(x_i) &\cong \hspace{-1mm}\int\hspace{-2mm} \prod_{j\in \partial a\setminus i}  \vcinfoidx{j}{a}{t+1}(x_i) \times e^{-\frac{\bracket{y_a - \sum_{j=1}^n A_{aj}x_j}^2}{2\sigma^2}}dx_j,
\end{align}
and the decoding step at each iteration becomes
\begin{align}
\label{eq:map_non_exp}
\hat{x}^{(t)}_i = \argmax_{x_i} \Prob(x_i|\by) \approx 
\argmax_{x_i}\wh{\prior}(x_i)\cdot \prod_{a\in \partial i} \cvinfoidx{a}{i}{t}(x_i).
\end{align}
Moreover, the 
functions
$h_{\mean}\bracket{\cdot; \cdot}$ and $h_{\var}\bracket{\cdot; \cdot}$ in \eqref{eq:de_equal_protect} are modified to 
$\wh{h}_{\mean}\bracket{\cdot; \cdot}$
and $\wh{h}_{\var}\bracket{\cdot; \cdot}$ as 
\[
 \wh{h}_{\mean}(\mu; v) =&\lim_{\gamma \rightarrow \infty} \frac{\int x_i \cdot e^{\gamma \log \wh{\prior}(x_i)}
    \cdot e^{-\frac{\gamma\bracket{x_i-\mu}^2}{2v} } dx_i}{
    \int  e^{\gamma \log \wh{\prior}(x_i)} \cdot 
    e^{-\frac{\gamma\bracket{x_i-\mu}^2}{2v} } dx_i}; \\
\wh{h}_{\var}(\mu; v) = & \lim_{\gamma \rightarrow \infty}\frac{\gamma\int x^2_i \cdot e^{\gamma \log \wh{\prior}(x_i)}
\cdot 
    e^{-\frac{\gamma\bracket{x_i-\mu}^2}{2v} } dx_i}{\int  e^{\gamma \log \wh{\prior}(x_i)}
\cdot     e^{-\frac{\gamma\bracket{x_i-\mu}^2}{2v} } dx_i} \\
-~& \bracket{\wh{h}_{\mean}(\mu; v)}^2.  
\]
Afterwards, we can design the 
sensing matrix with the same procedure as in \eqref{eqn:density_evolve_optim} and \eqref{eq:unequal_protect_optim_obj}.

\subsection{MMSE decoder}
Notice that both \eqref{eq:map_exp} and \eqref{eq:map_non_exp} 
reconstruct the signal by minimizing the error probability 
$\Prob\bracket{\hat{\bx} \neq \bx^{\natural}}$, which 
can be regarded as a MAP decoder. 
This subsection considers MMSE decoder, 
which is to minimize the $\ell_2$ error, i.e., 
$\norm{\hat{\bx} - \bx^{\natural}}{2}$. 
The message-passing procedure stays the same as 
\eqref{eq:general_message_pass} while the decoding procedure 
needs to be modified to

{\small \vspace{-3mm}
\begin{align*}
\hat{x}^{(t)}_i = \int x_i \Prob(x_i|\by)dx_i \approx 
\int \bracket{x_i \cdot \wh{\prior}(x_i)\cdot \prod_{a\in \partial i} \cvinfoidx{a}{i}{t}(x_i)}dx_i. 
\end{align*} \vspace{-2mm}
}

\par \noindent
Moreover, the functions 
$h_{\mean}(\cdot; \cdot)$ and $h_{\var}\bracket{\cdot; \cdot}$ 
in the DE in \eqref{eq:de_equal_protect}
are modified to $\wt{h}_{mean}\bracket{\cdot; \cdot}$
and $\wt{h}_{\var}\bracket{\cdot; \cdot}$ 
as  
\[ 
&\wt{h}_{\mean}(\mu; v) \hspace{-1mm}=\hspace{-1mm}  
\frac{\int x_i \cdot \wh{\prior}(x_i) \cdot   e^{-\frac{\bracket{x_i-\mu}^2}{2v} } dx_i}{
    \int  \wh{\prior}(x_i) \cdot    
    e^{-\frac{\bracket{x_i-\mu}^2}{2v} } dx_i}; \\
&\wt{h}_{\var}(\mu; v) \hspace{-1mm}= \hspace{-1mm}\frac{\int x^2_i\cdot
\wh{\prior}(x_i) \cdot e^{-\frac{\bracket{x_i-\mu}^2}{2v} } dx_i}{\int \wh{\prior}(x_i) \cdot   
    e^{-\frac{\bracket{x_i-\mu}^2}{2v} } dx_i}- \bracket{\wt{h}_{\mean}(\mu; v)}^2. 
\]
Having discussed two potential directions of generalization, next 
we will present the numerical experiments.

\section{Numerical Experiments}
\label{sec:simul}
This section presents the numerical experiments using both synthetic data and real-world data. 
We consider the sparse
signal and  compare the design of
preferential sensing with that of 
the regular sensing.
For the simplicity of the code design and
the construction of the corresponding sensing 
matrix,  we fix the degrees $\{\rho_{H, i}\}$
and  $\{\rho_{L, i}\}$ of the check nodes 
to $\rho_{H, \deg{c}_{\mathsf{H}}} = 1$
and $\rho_{L,\deg{c}_{\mathsf{L}}} = 1$, respectively.
Therefore, each check node
has $\deg{c}_{\mathsf{H}}$ edges connecting to 
the high-priority part $\bx^{\natural}_H$ and 
$\deg{c}_{\mathsf{L}}$ edges connecting to
 the low-priority 
part $\bx^{\natural}_L$. 
Then we construct the sensing matrix with the 
algorithm being illustrated in Alg.~\ref{alg:irregular_design}.

We evaluate two types of sensing matrices for the preferential sensing, 
namely, $\bA^{(\textup{init})}_{\textup{preferential}}$ and 
$\bA^{(\textup{final})}_{\textup{preferential}}$, which 
correspond to the distributions $\{\blambda_H\}$ and $\{\blambda_L \}$
in the initialization phase and at the final outcome of Alg.~\ref{alg:irregular_design}. 
As the baseline, we design the sensing matrix $\bA_{\textup{regular}}$
via \eqref{eqn:density_evolve_optim} 
which provides regular sensing with
an additional constraint which enforces equal edge number
with $\bA^{(\textup{final})}_{\textup{preferential}}$
 for a fair comparison.
 
\begin{algorithm}[h]
\caption{Design of Sensing Matrix for Preferential Sensing.}
\label{alg:irregular_design}
\begin{algorithmic}[1]
\Statex \textbullet~
\textbf{Input}: maximum variable node degree $\deg{v}_{\mathsf{max}}$, check node
degree $\deg{c}_{\mathsf{H}}$ and $\deg{c}_{\mathsf{L}}$, signal lengths $n_H$ and $n_L$, sparsity numbers $k_H$ and $k_L$, 
and iteration number $T$. 
\Statex \textbullet~
\textbf{Initialization}:
set $\beta_H \asymp \log\bracket{\frac{n_H}{k_L}}, 
\beta_L \asymp \log\bracket{\frac{n_L}{k_L}}$. 
Then we initialize $\{\lambda_{H, i}\}$ and
$\{\lambda_{L, i}\}$ as 
\begin{align*}
& \min_{\substack{\blambda_{H} \in \Delta_{\deg{v}_{\mathsf{max}} - 1},\\ \blambda_{L} \in \Delta_{\deg{v}_{\mathsf{max}}-1}}}
\sum_i i \lambda_{H, i}, \\
~\St~&
n_H \deg{c}_{\mathsf{L}}\bracket{\sum_i i\lambda_{H, i} } = 
n_L \deg{c}_{\mathsf{H}}\bracket{\sum_i i\lambda_{L, i} }; \\
& \bracket{\frac{\beta_H k_H}{n_H} 
\sum_{\ell}\frac{\lambda_{H, \ell}}{\ell}}^2 + 
\bracket{
\frac{\beta_L k_L}{n_L}
\sum_{\ell}\frac{\lambda_{L, \ell}}{\ell}}^2\\
\leq & \quad  \frac{1}{(\deg{c}_{\mathsf{H}})^2 + (\deg{c}_{\mathsf{L}})^2}; \\
& \sum_{\ell}\frac{\lambda_{H, \ell}}{\sqrt{\ell}}\leq \frac{\sqrt{n_H}}{\sqrt{k_H}\sqrt{\deg{c}_{\mathsf{H}} + \deg{c}_{\mathsf{L}}}}; \\
& \lambda_{L, 1} = \lambda_{H, 1} =0,  
\end{align*}
which is equivalent to 
\eqref{eq:unequal_protect_optim_obj} 
without the Requirement \ref{require:noisy_unequal}.

\Statex \textbullet~
\textbf{Iterative Update:}
denote $\blambda^{(t)}_H$ (or $\blambda^{(t)}_L$) as the 
updated version of $\blambda_{(H)}$ (or $\blambda_{(L)}$) at the 
$t$th iteration.    

\Statex \quad \textbullet~\textbf{For time $t =1$ to $T$:}
update $\blambda^{(t)}_{H}$ and $\blambda^{(t)}_L$ 
by alternating minimization of 
\eqref{eq:unequal_protect_optim_obj} with 
Requirement \ref{require:noiseless_unequal} and 
Requirement
\ref{require:noisy_unequal} being 
replaced by 
\eqref{eq:unequal_variance}, \eqref{eq:unequal_error}, and \eqref{eq:unequal_error_decrease_faster}. 

\begin{enumerate}

\item 
\textbf{Update $\blambda^{(t)}_H$} 
with  $\blambda_L$ being fixed to be 
$\blambda^{(t-1)}_L$;

\item 
\textbf{Update $\blambda^{(t)}_L$} 
with $\blambda_{H}$ being fixed to be 
$\blambda^{(t)}_H$.
\end{enumerate}
\Statex \textbullet~
\textbf{Output}: degree distribution $\blambda^{(T)}_H$ and $\blambda^{(T)}_L$.

\end{algorithmic}
\end{algorithm}

\subsection{Experiments with synthetic data}


\noindent{\textbf{Experiment set-up}.}
We fix the check node degrees 
$\deg{c}_{\mathsf{H}}$ and $\deg{c}_{\mathsf{L}}$ as $5$ and let 
the maximum variable node degree $\deg{v}_{\mathsf{max}}$
as 
$50$.   
The magnitude of the 
non-zero entries is set to $1$. Then we 
study the recovery performance with varying $\snr = \norm{\bx^{\natural}}{2}^2/\sigma^2$. 
The following numerical experiments 
separately study 
the impact of the signal length 
$n_H$ and $n_L$
and the impact of the sparsity number 
$k_H$ and $k_L$.

\subsubsection{Impact of sparsity number}
\begin{figure}[h]
\centering
\vspace{2mm}
\mbox{
\includegraphics[width = 1.7in]{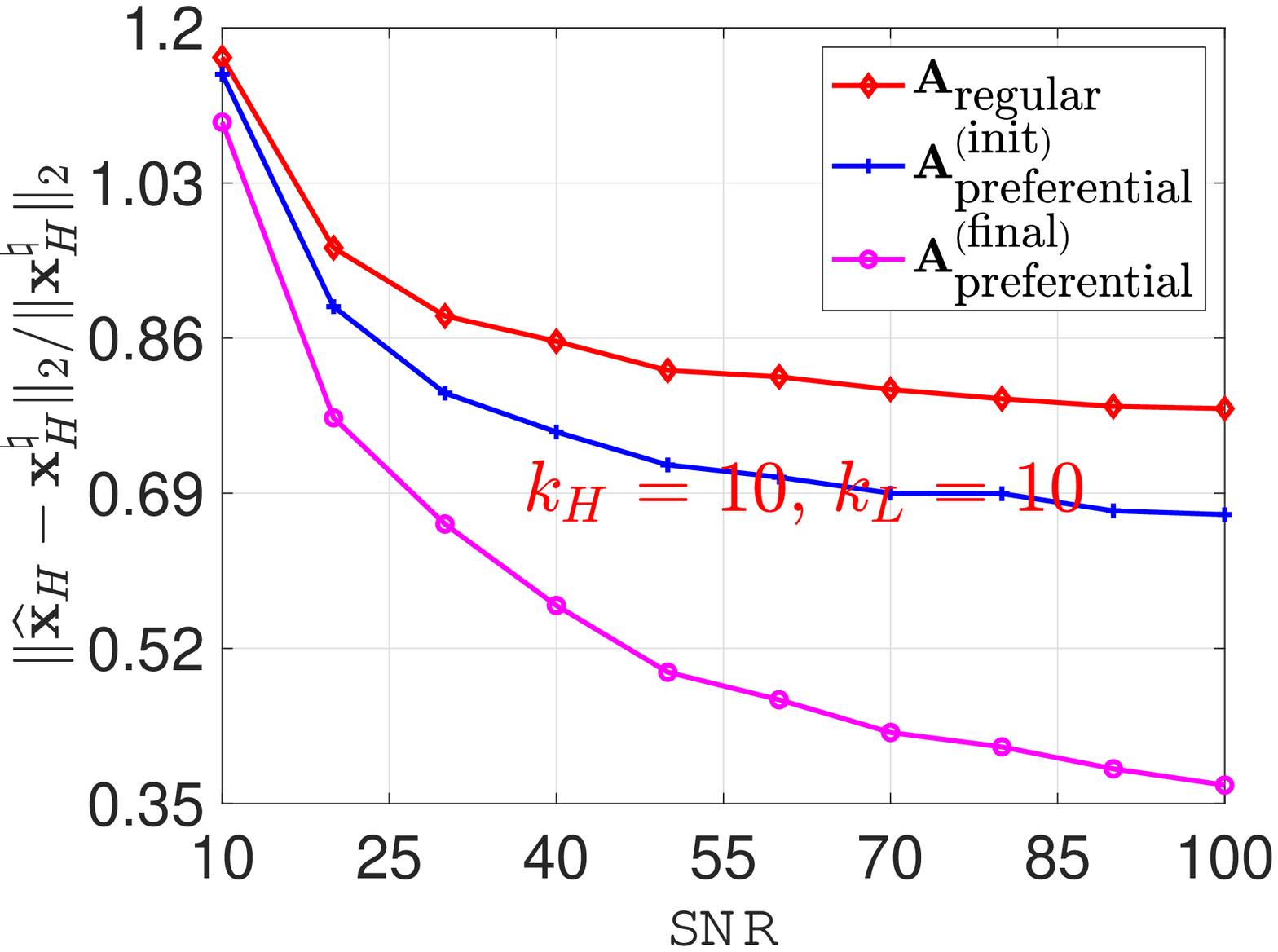}
\includegraphics[width = 1.7in]{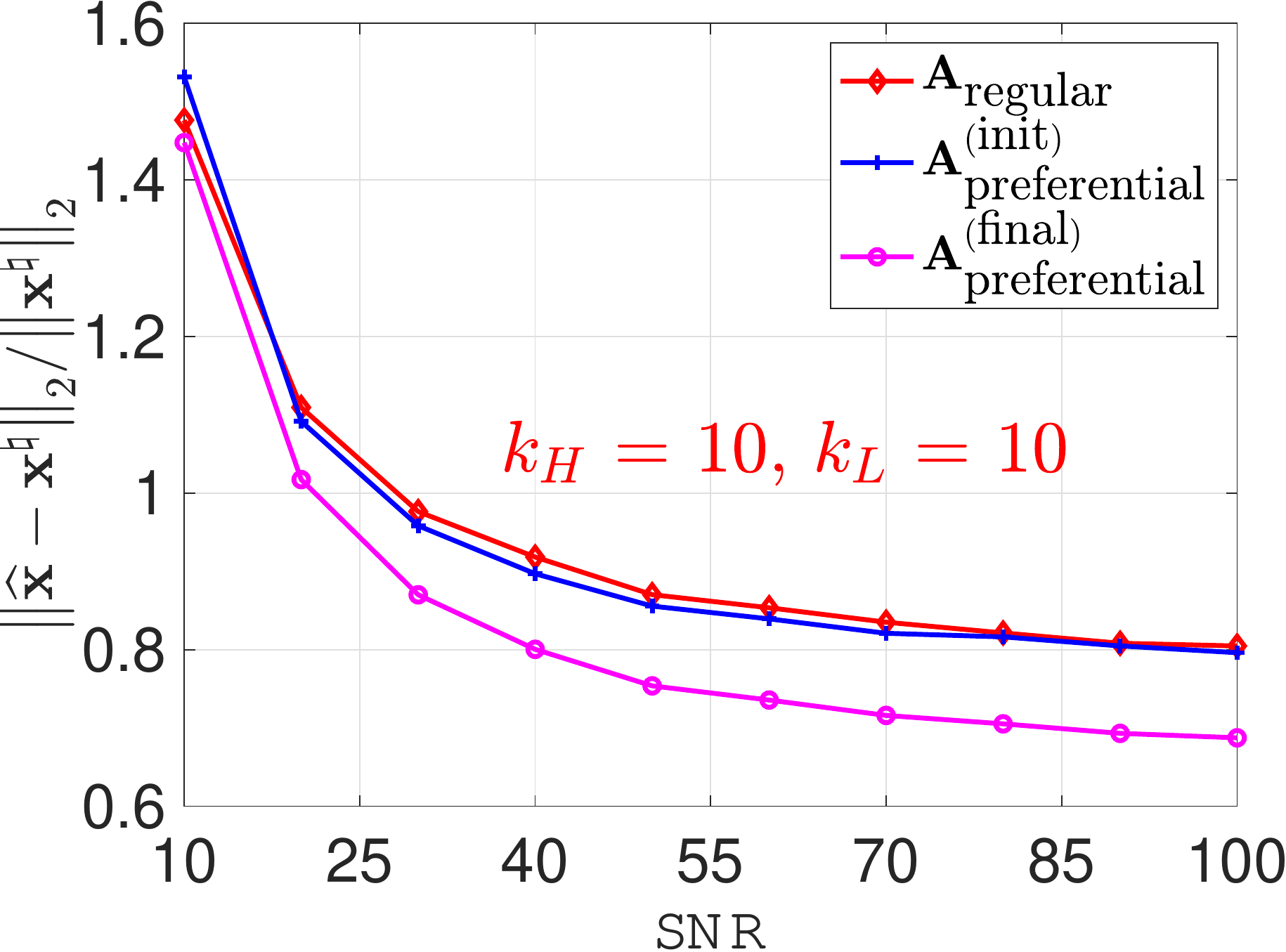}
}

\mbox{
\includegraphics[width = 1.7in]{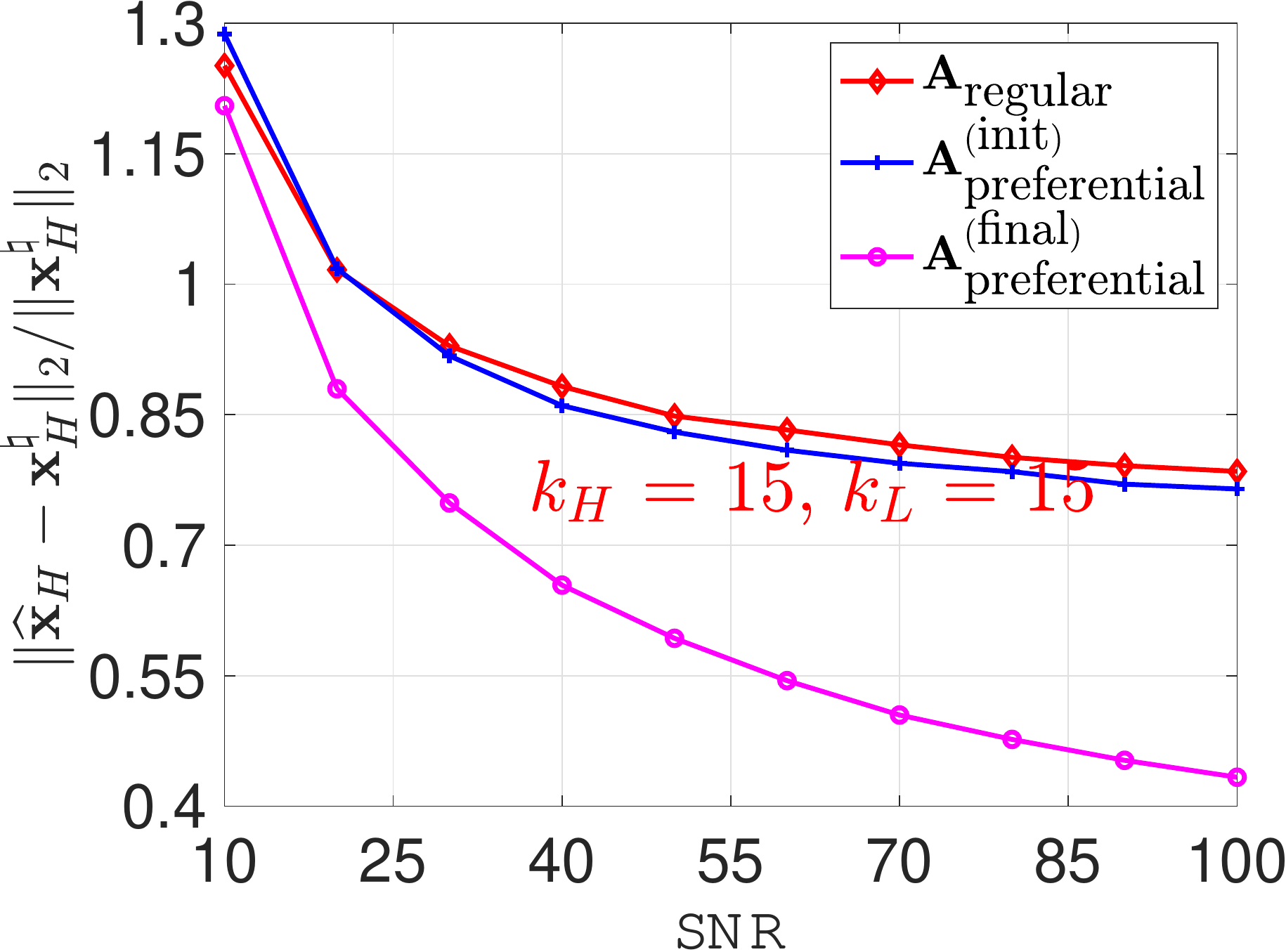}
\includegraphics[width = 1.7in]{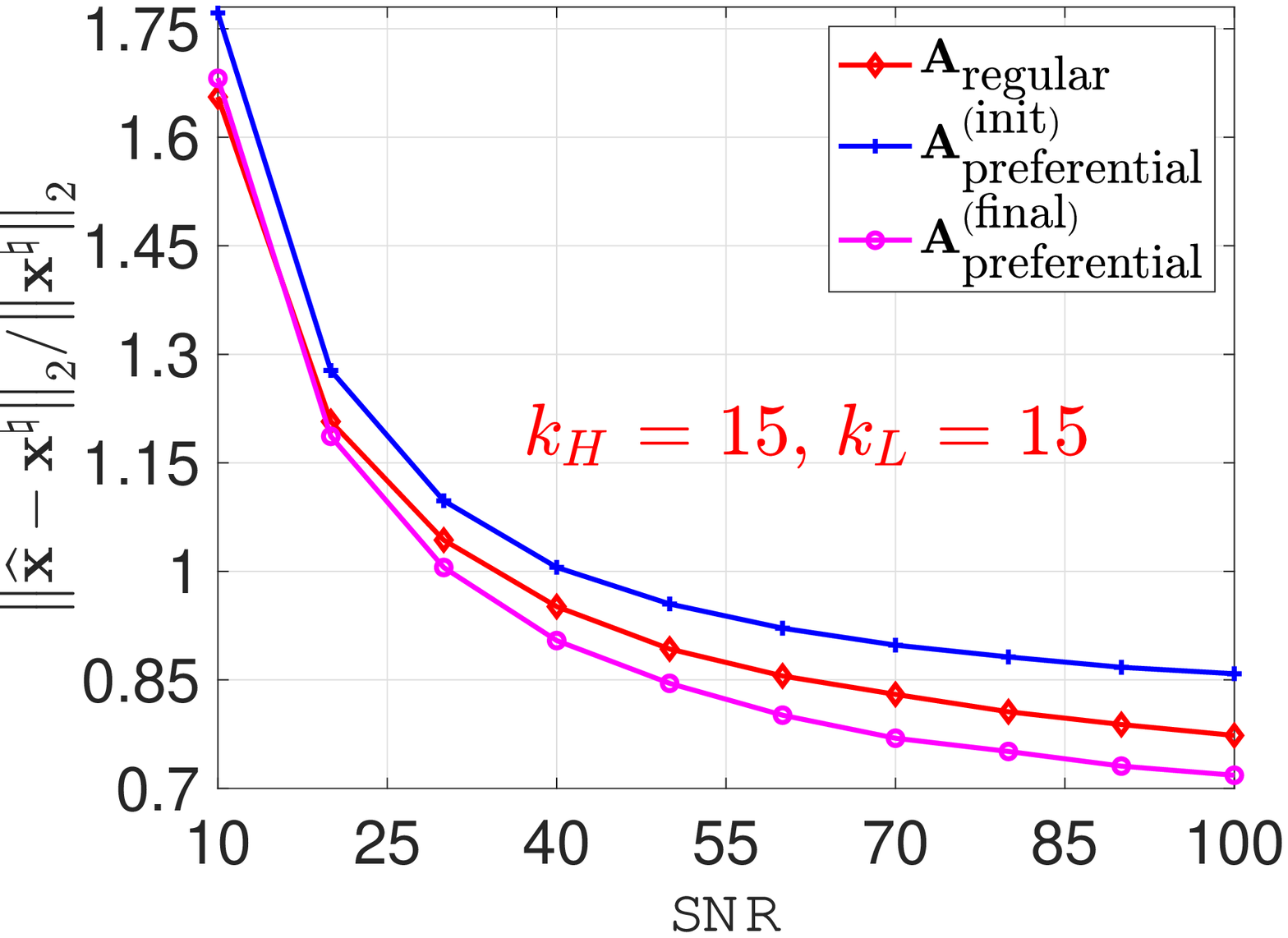}
}

\mbox{
\includegraphics[width = 1.7in]{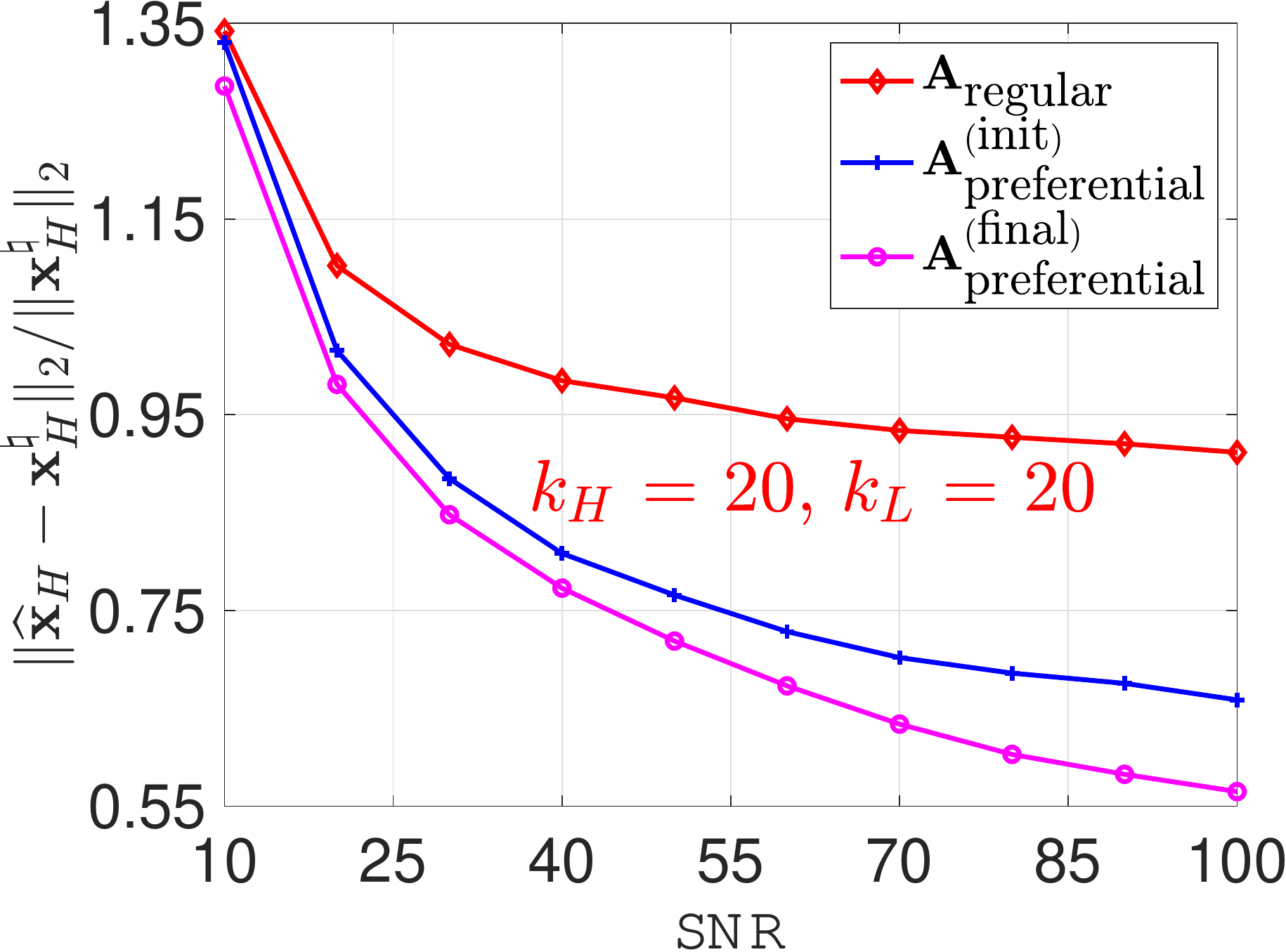}
\includegraphics[width = 1.7in]{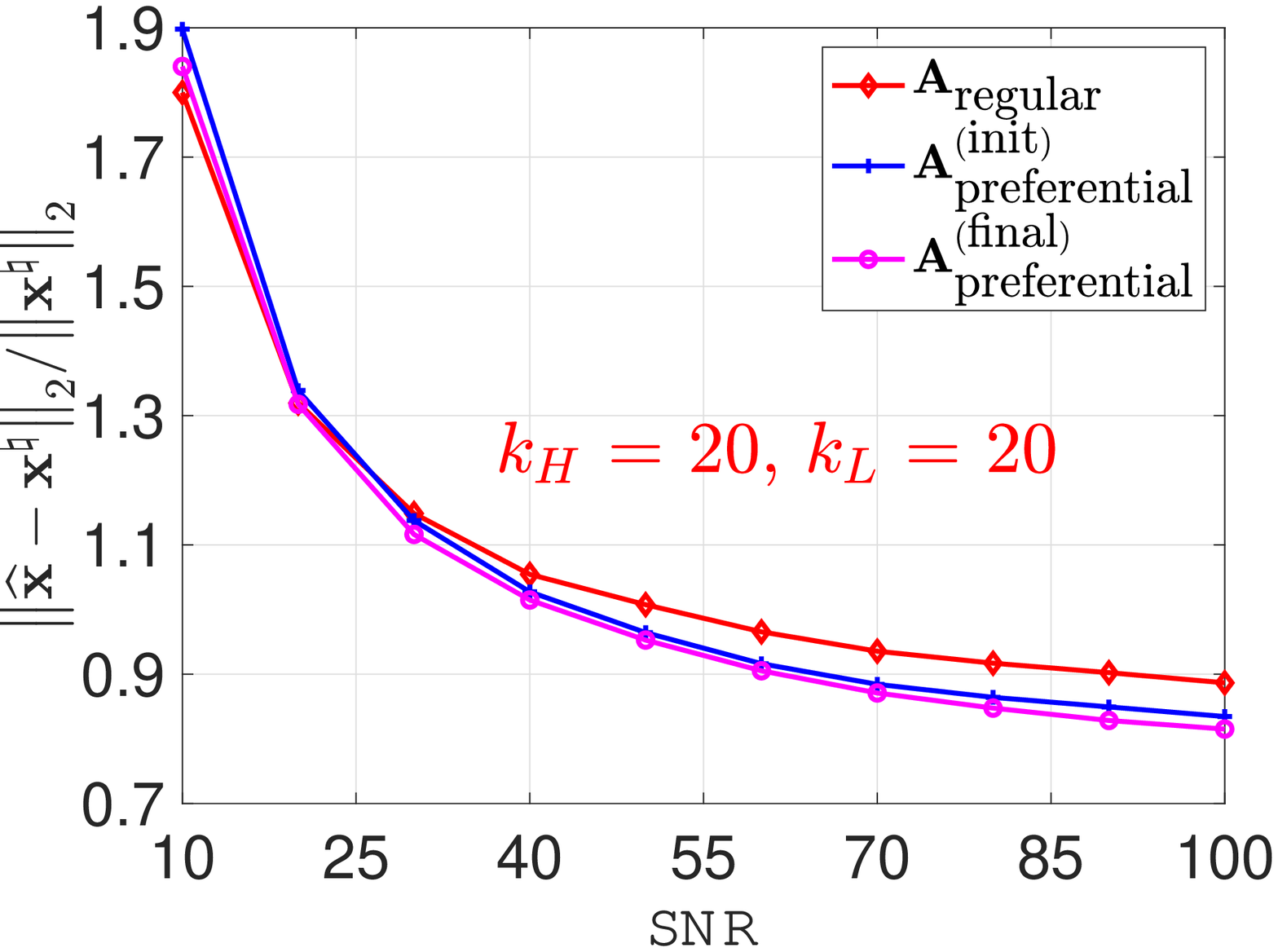}
}

\caption{Comparison of 
preferential sensing vs regular sensing. 
The length $n_H$ of the high-priority part 
$\bx^{\natural}_H$ is set as $100$; while the 
length $n_L$ of the low-priority part 
$\bx^{\natural}_L$ is set as $400$. 
(\textbf{Left panel}) We evaluate the 
reconstruction performance w.r.t. 
the high-priority part $\|\wh{\bx}_{H} - \bx^{\natural}_H\|_{2}/\|\bx^{\natural}_H\|_{2}$. 
(\textbf{Right panel}) We evaluate the 
reconstruction performance w.r.t. 
the whole signal $\|\wh{\bx} - \bx^{\natural}\|_{2}/\|\bx^{\natural}\|_{2}$. 
}
\label{fig:kfactor}
\end{figure}

We fix the length $n_H$ of the high-priority 
part $\bx_H^{\natural}$ as  
$100$ and the length $n_L$ of the low-priority 
part $\bx_L^{\natural}$ as  $400$. 
The simulation results are plotted in 
Fig.~\ref{fig:kfactor}.

We first investigate the recovery performance w.r.t. the 
high priority part $\bx^{\natural}_H$. 
Using the sensing matrix $\bA_{\textup{regular}}$
(regular sensing) as the baseline, we conclude that 
our sensing matrix $\bA^{(\textup{final})}_{\textup{preferential}}$ 
(preferential sensing) achieves better performance 
when the signal is more sparse. Consider 
the case when $\snr = 100$. When $k_H = k_L = 10$, 
the ratio $\|\wh{\bx}_H - \bx^{\natural}_H\|_2/\|\bx^{\natural}_H\|_2$ for $\bA^{(\textup{final})}_{\textup{preferential}}$ is approximately $0.35$ while that 
of the $\bA_{\textup{regular}}$ is $0.86$. 
When the sparsity number $k_H$ and $k_L$ increase to $15$, the 
improvement is approximately $(0.85-0.4)/0.85 \approx 53\%$. 
When the sparsity number $k_H$ and $k_L$ increase to $20$, 
the corresponding improvement further decreases to $(0.95-0.55)/0.95 \approx 42\%$.

When turning to the reconstruction error $\norm{\wh{\bx} - \bx^{\natural}}{2}/\norm{\bx^{\natural}}{2}$ w.r.t. the 
whole signal, we notice a similar phenomenon, i.e., a sparser signal 
contributes to better performance. 
Additionally, we notice the sensing matrix $\bA_{\textup{preferential}}^{(\textup{final})}$ achieves significant improvements 
in comparison to its initialized version 
$\bA^{(\textup{init})}_{\textup{preferential}}$.

\subsubsection{Impact of signal length} 
\begin{figure}[h]
\centering
\vspace{2mm}
\mbox{
\includegraphics[width = 1.7in]{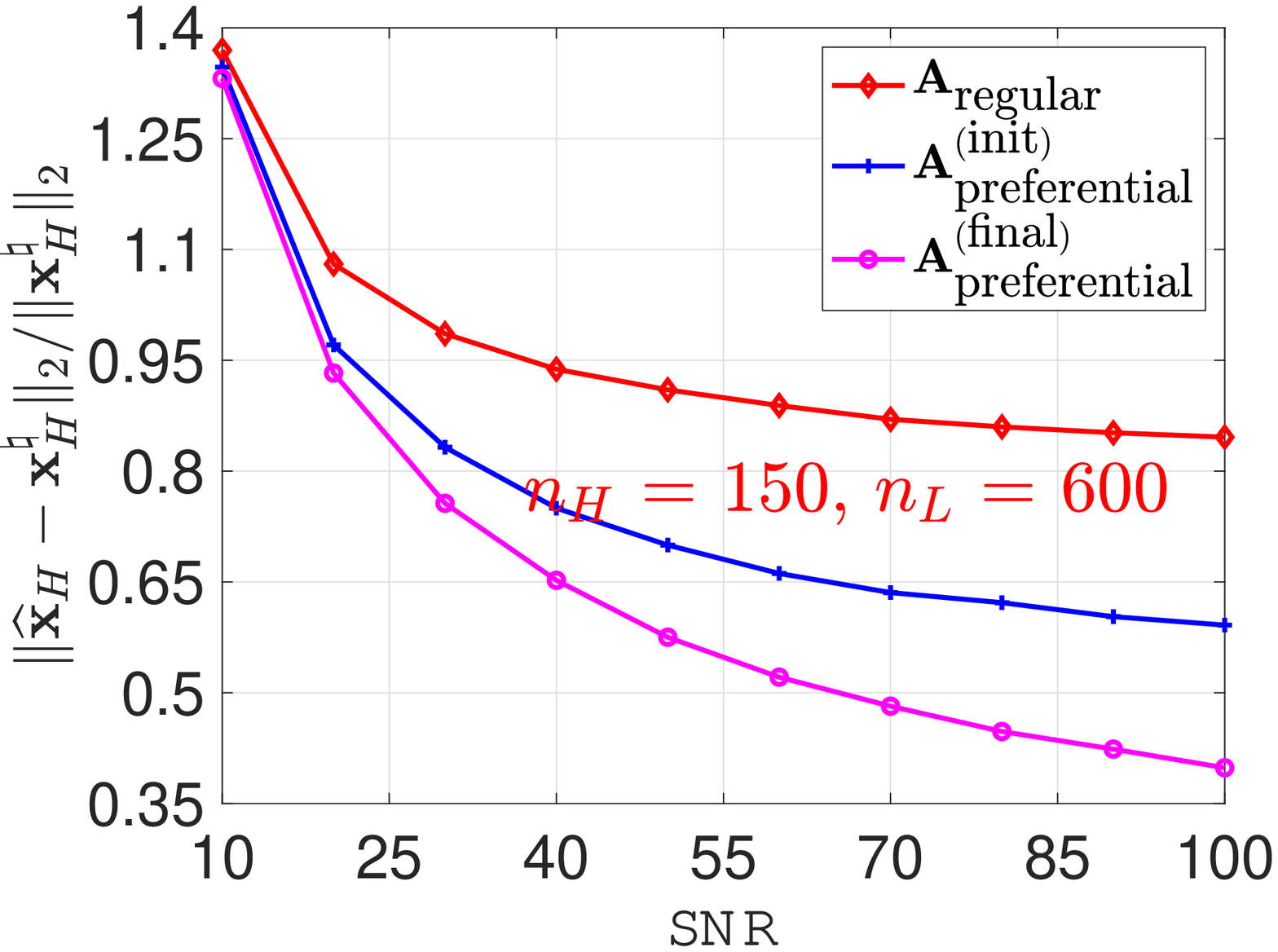}
\includegraphics[width = 1.7in]{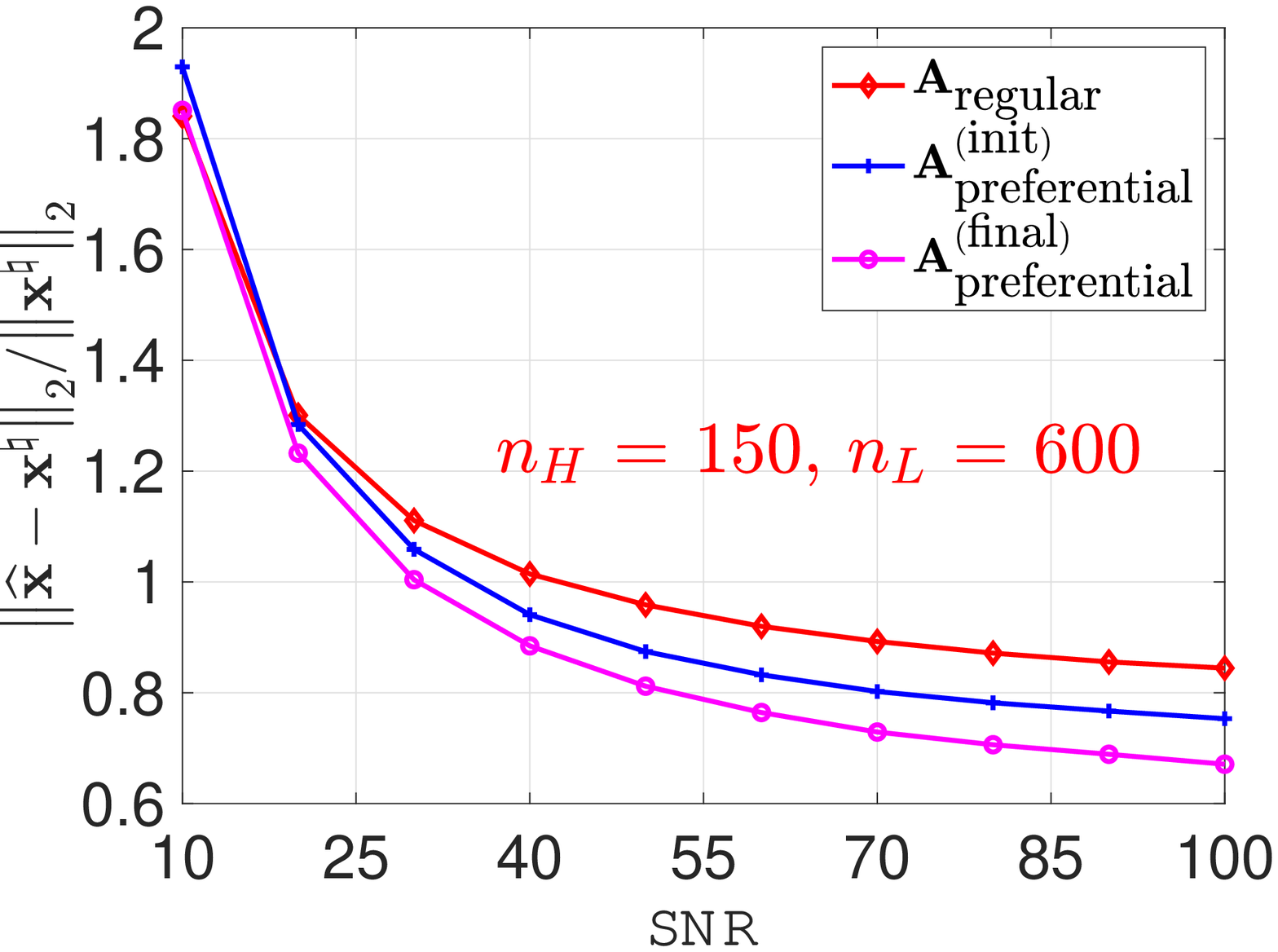}
}

\mbox{
\includegraphics[width = 1.7in]{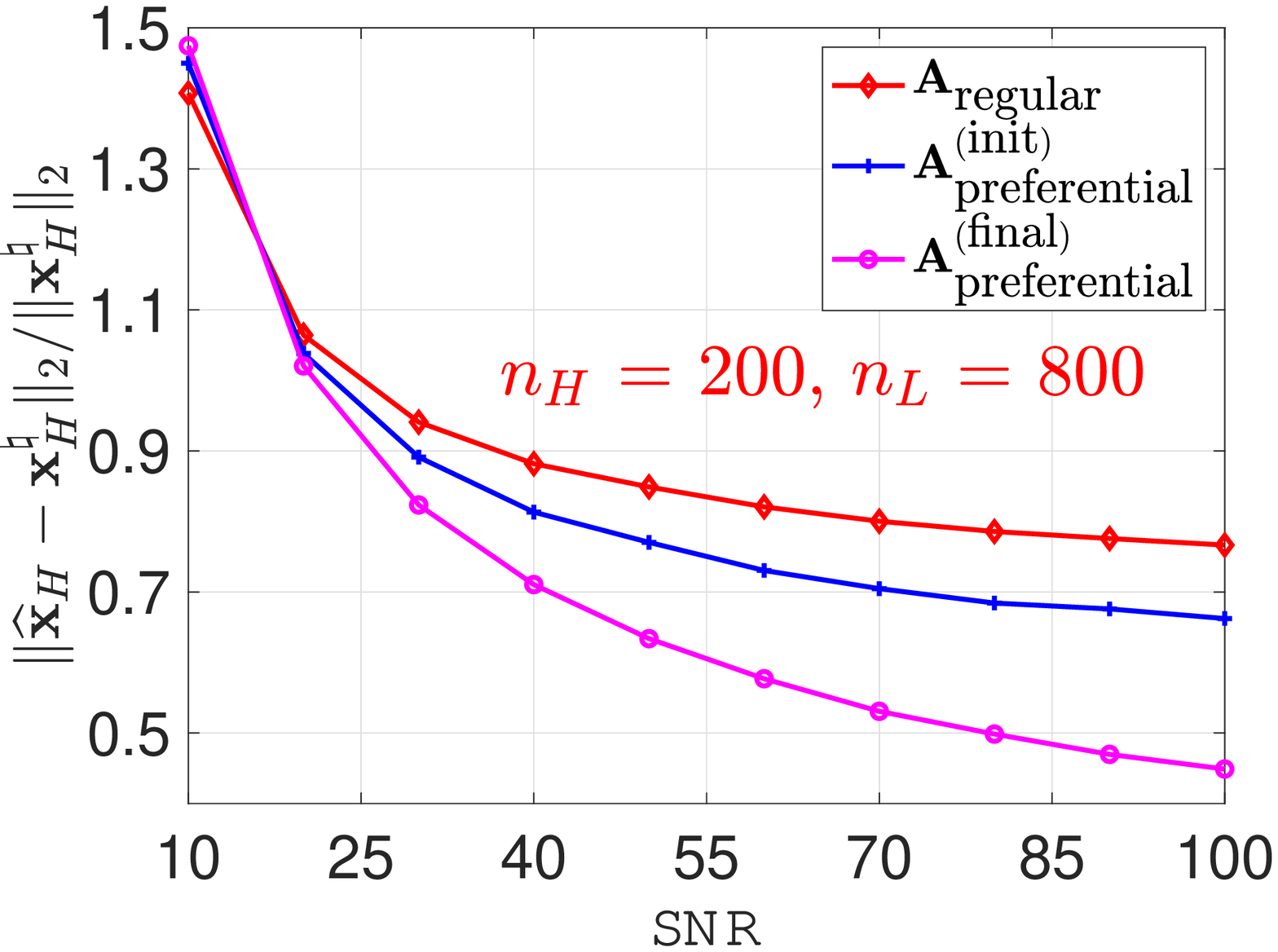}
\includegraphics[width = 1.7in]{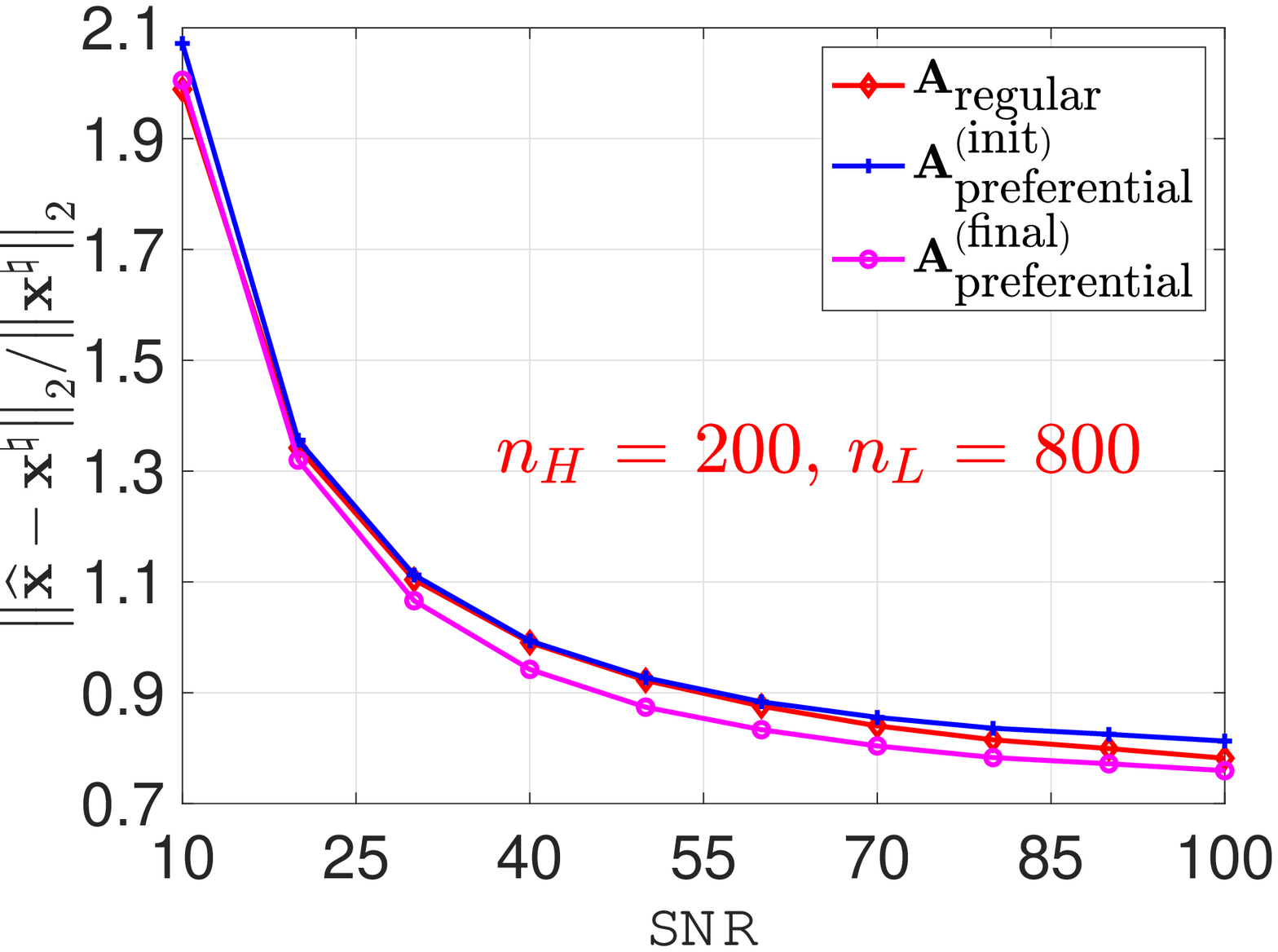}
}

\mbox{
\includegraphics[width = 1.7in]{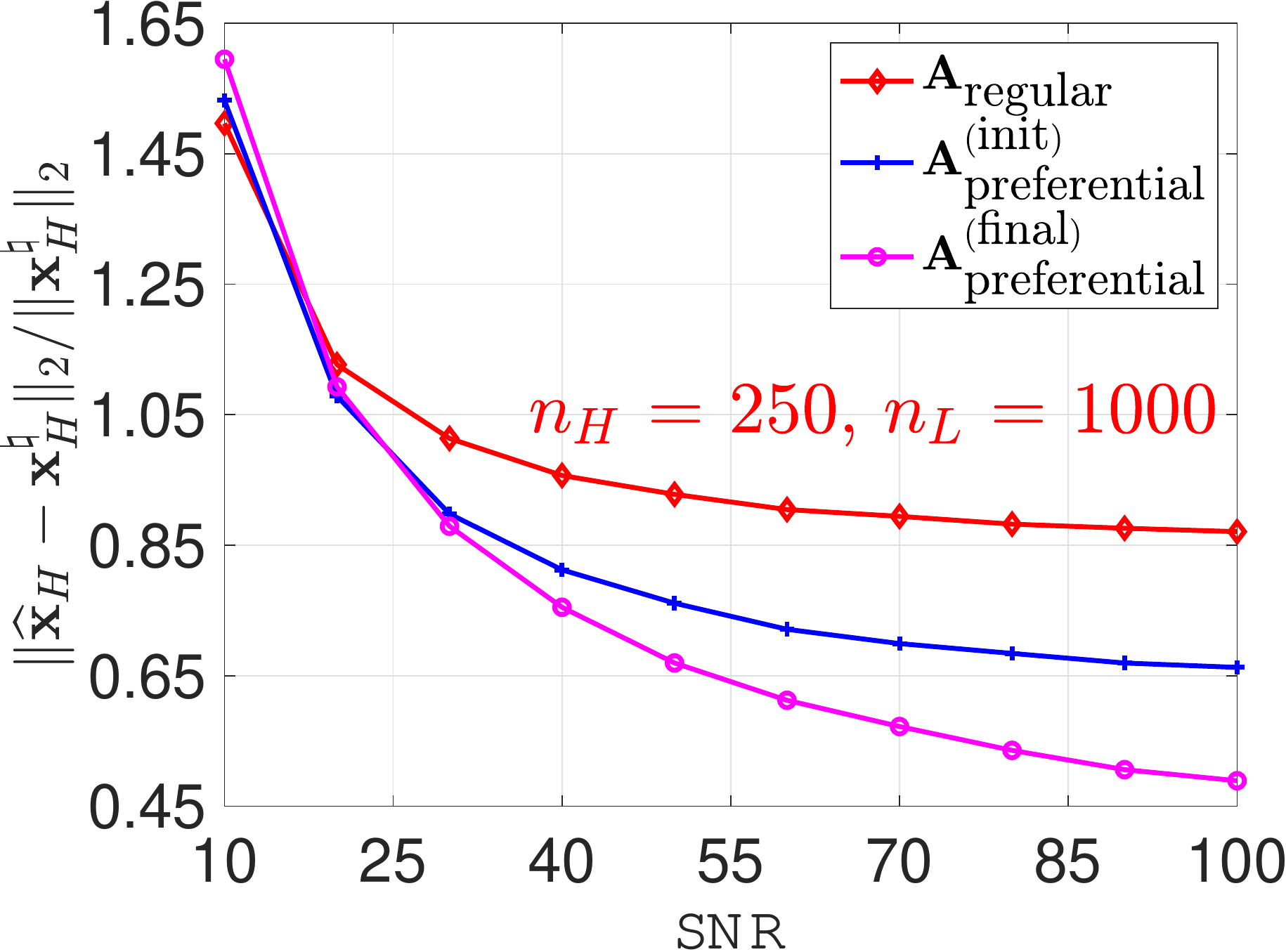}
\includegraphics[width = 1.7in]{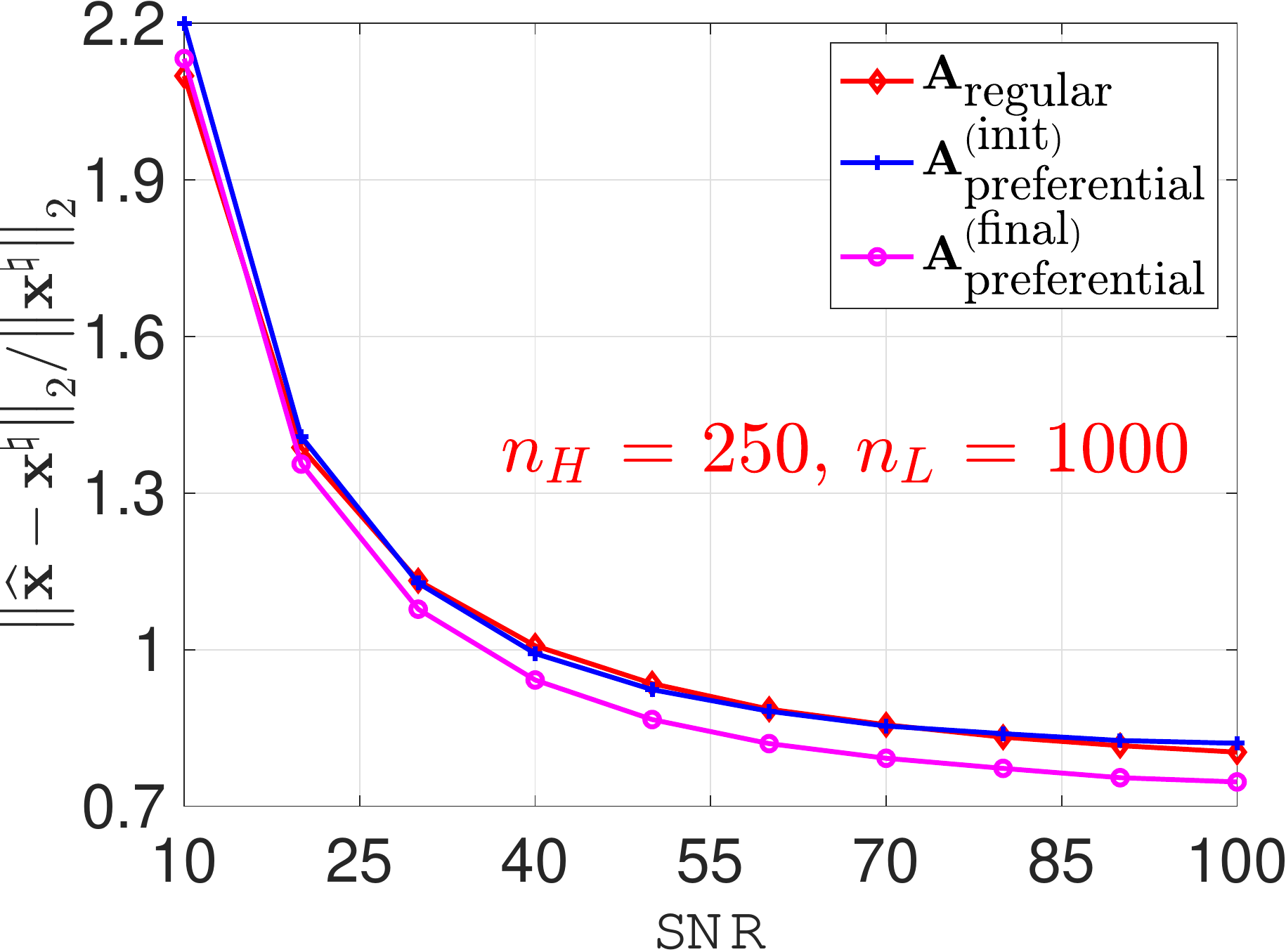}
}

\mbox{
\includegraphics[width = 1.7in]{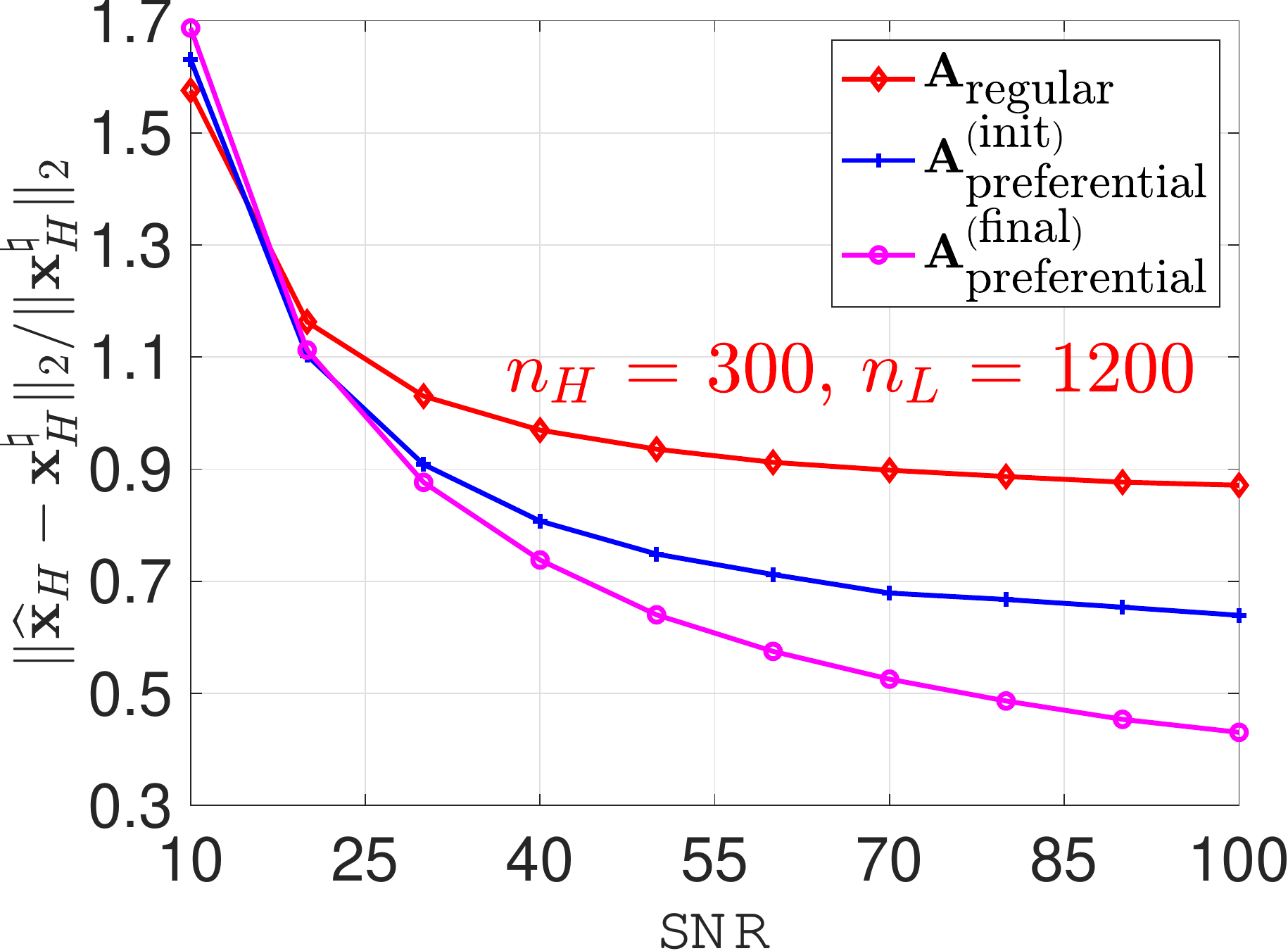}
\includegraphics[width = 1.7in]{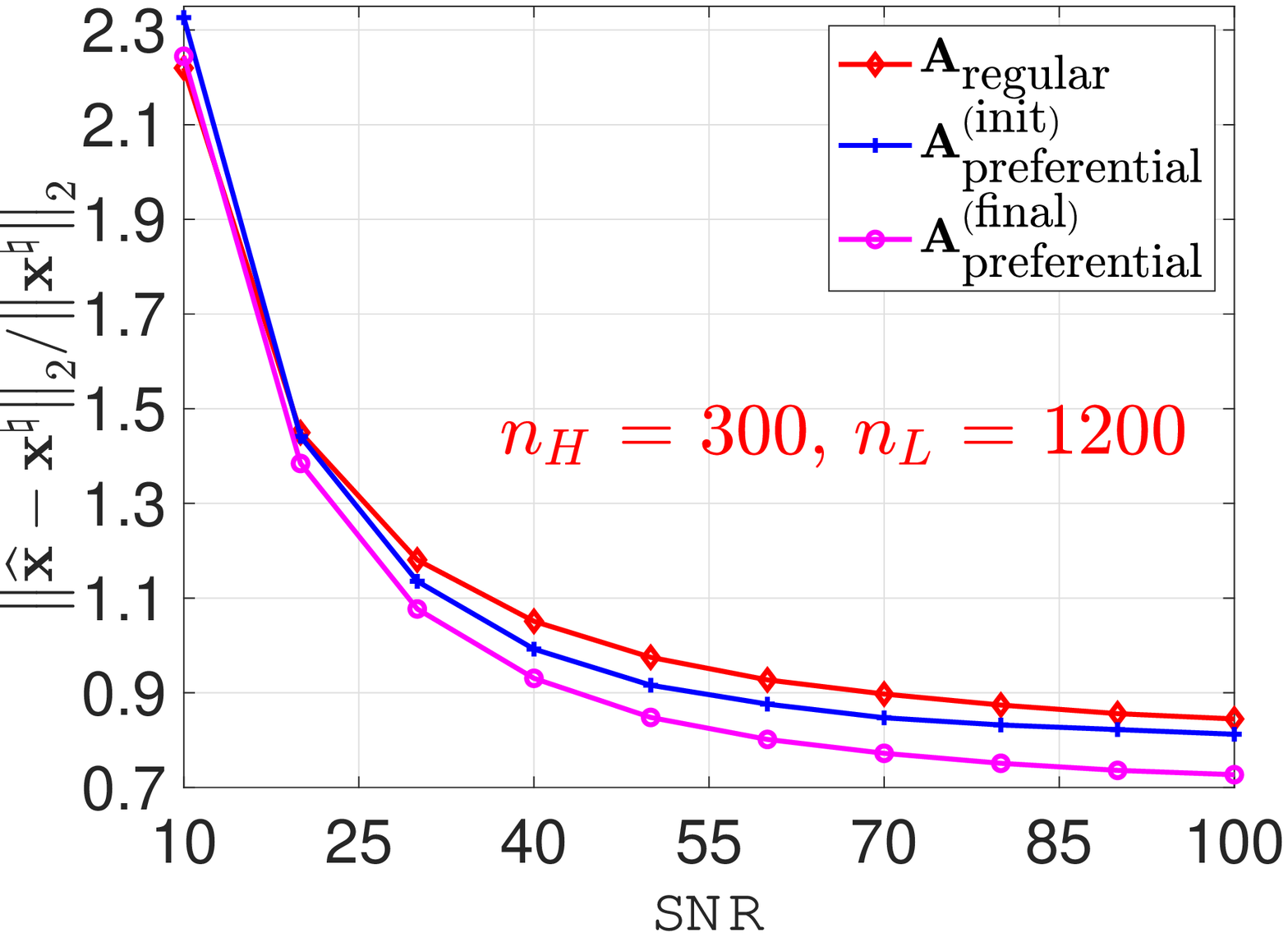}
}

\caption{Comparison of 
preferential sensing vs regular sensing. 
Both the sparsity number $k_H$ and 
$k_L$ are set as $15$. 
(\textbf{Left panel}) We evaluate the 
reconstruction performance w.r.t.
the high-priority part $\|\wh{\bx}_{H} - \bx^{\natural}_H\|_{2}/\|\bx^{\natural}_H\|_{2}$. 
(\textbf{Right panel}) We evaluate the 
reconstruction performance w.r.t.
the whole signal $\|\wh{\bx} - \bx^{\natural}\|_{2}/\|\bx^{\natural}\|_{2}$. 
}

\label{fig:nfactor_k15}
\end{figure}

We also studied 
various settings in which the length $n_H$  of the 
high-priority part $\bx^{\natural}_H$ is set to $\set{150, 200, 250, 300}$ and the corresponding length $n_L$ of the 
low-priority part $\bx^{\natural}_L$ is set to 
$\set{600, 800, 1000, 1200}$. 
The simulation results are 
plotted in Fig.~\ref{fig:nfactor_k15}.

Compared to regular sensing, our sensing matrix 
$\bA^{(\textup{final})}_{\textup{preferential}}$
can reduce the error
 in the high-priority part $\bx^{\natural}_H$ significantly.  
For example, when $\snr = 100$, 
the ratio $\|\wh{\bx}_H - \bx^{\natural}_H\|_2/\|\bx_H^{\natural}\|_2$ reduces between $40\%\sim 60\%$ with the 
sensing matrix $\bA^{(\textup{final})}_{\textup{preferential}}$. 
Meanwhile, w.r.t. the whole signal $\bx^{\natural}$, the ratio $\norm{\wh{\bx} - \bx^{\natural}}{2}/\norm{\bx^{\natural}}{2}$ decreases
with a smaller magnitude.

\subsection{Experiments with real-world data}
We compare the performance of sensing matrices
for images using
$(i)$  MNIST dataset
\cite{lecun1998gradient}, which consists of 
$10000$ images in the testing set 
and $60000$ images in the training set; and
$(ii)$ Lena image. 

To obtain a sparse representation for each image, 
we perform a $2\textup{D}$ Haar transform $\calH(\cdot)$, 
which generates four sub-matrices being called 
as the approximation coefficients (at the coarsest level), 
horizontal detail coefficients, vertical detail coefficients, and diagonal detail 
coefficients.  
The approximation coefficients are at the coarsest level 
and are treated as the high-priority part 
$\bx^{\natural}_H$; while the horizontal detail coefficients, vertical detail 
coefficients, 
and diagonal detail coefficients are regarded as the low-priority part 
$\bx^{\natural}_L$.
Hence we can write the sensing relation in 
\eqref{eq:sense_relation}
as 
\begin{align}
\label{eq:haar_sense_relation}
\by = \bA \calH(\textup{Image}) + \bw, 
\end{align}
where $\textup{Image}$ denotes the  input image,
$\calH(\cdot)$ denotes the vectorized version of the coefficients
and is viewed  as the sparse ground-truth signal, and 
$\bw$ denotes the sensing noise. 
The 
sensing matrix 
$\bA$ is designed such that the approximation coefficients of 
$\calH(\textup{Image})$ can be better reconstructed. 

\begin{figure*}
\centering
\includegraphics[width = 6.0in]{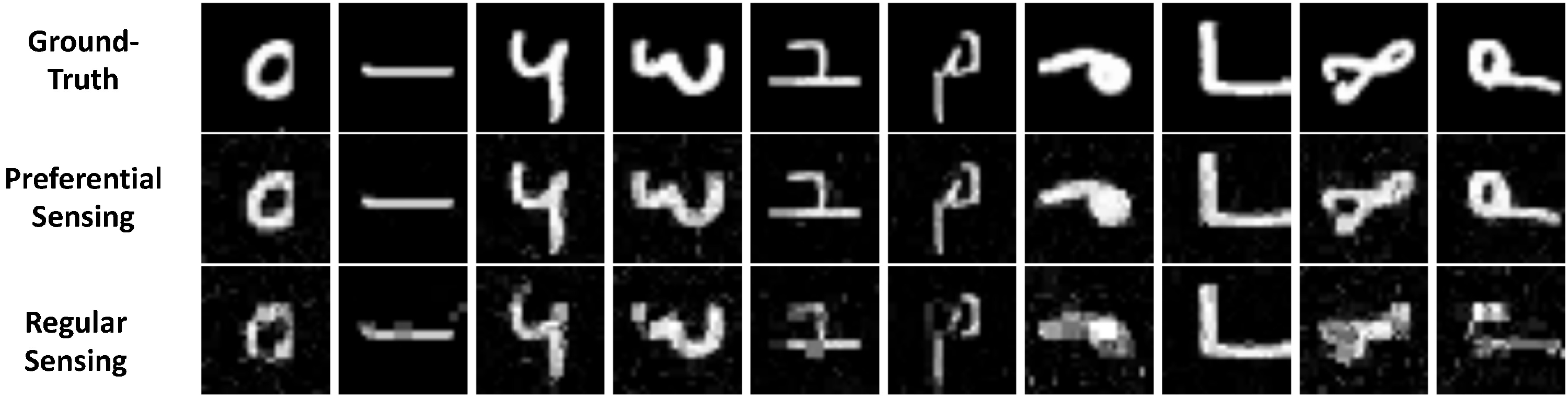}
\caption{The performance comparison between the 
sensing matrix for preferential sensing $\bA^{(\textup{final})}_{\textup{preferential}}$ and 
sensing matrix for regular sensing 
 $\bA_{\textup{regular}}$. 
 (\textbf{Top}) The ground-truth images. 
 (\textbf{Middle}) The reconstructed images with the 
 sensing matrix $\bA^{(\textup{final})}_{\textup{preferential}}$. 
 (\textbf{Bottom}) The reconstructed images with the 
 sensing matrix $\bA_{\textup{regular}}$. 
}
\label{fig:mnist_comp}
\end{figure*}

\subsection{Experiments with MNIST}

\noindent\textbf{Experiment set-up}.
We set the images from MNIST as the input, which 
consists of $10000$ images in the testing set 
and $60000$ images in the training set with 
each image being of dimension $28\times 28$. 

The whole datasets can be divided into $10$ categories
with each category representing a digit from zero to nine.
For each digit, we design one unique sensing matrix.
The lengths $n_H$ and $n_L$ are set to $(28/2)^2 = 196$ and $3 \times (28/2)^2 = 588$, respectively.
The sparsity coefficients $k_H$ and $k_L$ varied 
among different digits.

\noindent\textbf{Discussion}.
To evaluate the performance, we define 
ratios $r_{H, (\cdot)}$ and $r_{W, (\cdot)}$ as 
\[
r_{H, (\cdot)} & \defequal \frac{\|\wh{\bx}_H - \bx^{\natural}_H\|_2}{\|\bx^{\natural}_H\|_2}; \\
r_{W, (\cdot)} & \defequal \frac{\|\wh{\bx} - \bx^{\natural}\|_2}{\|\bx^{\natural}\|_2},
\]
which correspond to the $\ell_2$ error in
the high-priority part $\bx^{\natural}_H$ and the
entire signal
$\bx^{\natural}$, respectively. 
We use the sensing matrix $\bA_{\textup{regular}}$ 
as the benchmark. 
In addition, we omit the results of $\bA^{(\textup{init})}_{\textup{preferential}}$,
since the sensing matrix $\bA^{(\textup{final})}_{\textup{preferential}}$
has better performance.

The results are listed in Tab.~\ref{tab:mnist_comp}.
A subset of the reconstructed images are shown
in Fig.~\ref{fig:mnist_comp}. 
From Tab.~\ref{tab:mnist_comp} and Fig.~\ref{fig:mnist_comp}, 
we conclude that our sensing matrix $\bA^{(\textup{final})}_{\textup{preferential}}$
for the preferential sensing 
can better preserve the images when comparing with the 
sensing matrix $\bA_{\textup{regular}}$ for the regular sensing.

{ 
\begin{table*}[t]
\centering
\begin{tabular}{@{}cccccrlccc@{}}\toprule
 & \multicolumn{4}{c}{Training Set} & \phantom{abcd}& \multicolumn{4}{c}{Testing Set} 
\\
\cmidrule{2-5} \cmidrule{7-10} 
\textbf{Digit} &$r_{H, (p)}$ & $r_{H, (r)}$ & $r_{W, (p)}$ & $r_{W, (r)}$ &&   
$r_{H, (p)}$ & $r_{H, (r)}$ & $r_{W, (p)}$ & $r_{W, (r)}$\\ \midrule
$0$&$\textbf{0.28315}$&$0.5154$&$\textbf{0.44818}$&$0.60131$&&$\textbf{0.30292}$&$0.45749$&$\textbf{0.46283}$&$0.56486$\\
$1$&$\textbf{0.16746}$&$0.33751$&$\textbf{0.29332}$&$0.41599$&&$\textbf{0.1511}$&$0.45264$&$\textbf{0.2659}$&$0.51864$\\
$2$&$\textbf{0.26303}$&$0.50365$&$\textbf{0.42984}$&$0.59959$&&$\textbf{0.24896}$&$0.4233$&$\textbf{0.42216}$&$0.52556$\\
$3$&$\textbf{0.24613}$&$0.43677$&$\textbf{0.42514}$&$0.53163$&&$\textbf{0.26446}$&$0.46766$&$\textbf{0.43534}$&$0.56189$\\
$4$&$\textbf{0.28331}$&$0.44377$&$\textbf{0.44623}$&$0.53791$&&$\textbf{0.30092}$&$0.4445$&$\textbf{0.45804}$&$0.53749$\\
$5$&$\textbf{0.28405}$&$0.53511$&$\textbf{0.45727}$&$0.6198$&&$\textbf{0.27258}$&$0.47044$&$\textbf{0.44382}$&$0.56622$\\
$6$&$\textbf{0.28801}$&$0.39436$&$\textbf{0.45053}$&$0.51701$&&$\textbf{0.27084}$&$0.5086$&$\textbf{0.44134}$&$0.59534$\\
$7$&$\textbf{0.25503}$&$0.41621$&$\textbf{0.41809}$&$0.52896$&&$\textbf{0.27266}$&$0.51329$&$\textbf{0.41693}$&$0.5783$\\
$8$&$\textbf{0.31263}$&$0.51918$&$\textbf{0.47618}$&$0.61492$&&$\textbf{0.32731}$&$0.48163$&$\textbf{0.48699}$&$0.5837$\\
$9$&$\textbf{0.30171}$&$0.54394$&$\textbf{0.45241}$&$0.61799$&&$\textbf{0.27385}$&$0.55313$&$\textbf{0.43116}$&$0.62785$\\
\bottomrule
\end{tabular}
\vspace{4mm}
\caption{
The index $i = p$ corresponds to the sensing matrix $\bA^{(\textup{final})}_{\textup{preferential}}$ for the preferential sensing; 
while the index $i = r$ corresponds to the sensing matrix $\bA_{\textup{regular}}$ for the regular sensing.
We define the ratio $r_{H, (i)}$ ($i = \set{p, r}$) as the error 
w.r.t. the high priority part, namely,  $\|\wh{\bx}_H -\bx^{\natural}_H\|_{2}/\|\bx^{\natural}_H\|_{2}$.
Similarly we define the ratio $r_{W, (i)}$ ($i = \set{p, r}$) as the 
ratio w.r.t. the whole signal, namely, 
$\|\wh{\bx} -\bx^{\natural}\|_{2}/\|\bx^{\natural}\|_{2}$.
Moreover, we put the results corresponding to the 
sensing matrix $\bA^{(\textup{final})}_{\textup{preferential}}$
in the bold font. 
 }
 \label{tab:mnist_comp}
\end{table*}
}

\begin{figure*}
\centering
\mbox{
\includegraphics[width = 1.9in]{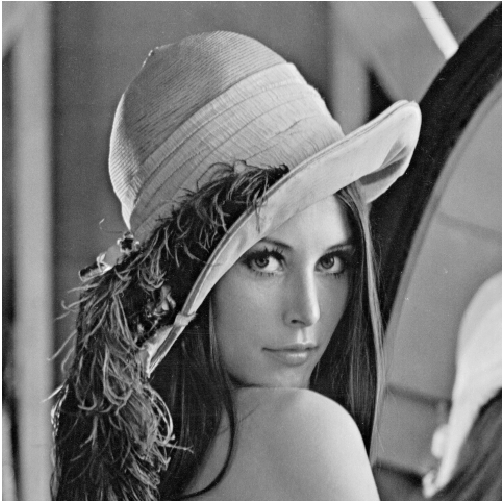}
\includegraphics[width = 1.9in]{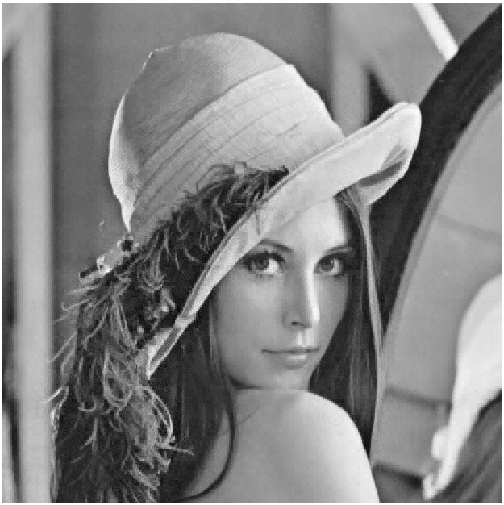}
\includegraphics[width = 1.9in]{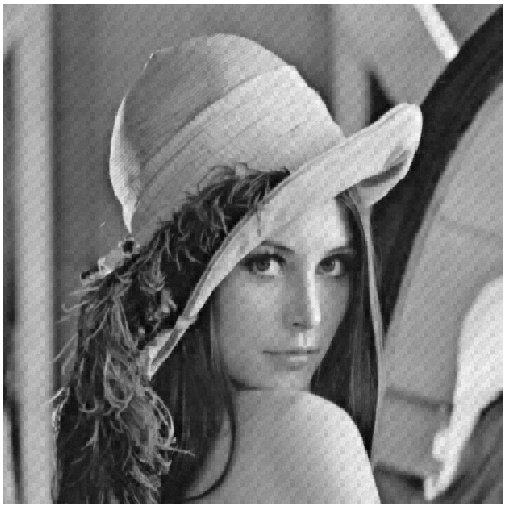}
}
\caption{(\textbf{Left}) Ground-truth image. 
(\textbf{Middle}) Reconstructed image via sensing matrix 
$\bA^{(\textup{final})}_{\textup{preferential}}$ for preferential sensing.
(\textbf{Right})  Reconstructed image via sensing matrix 
$\bA_{\textup{regular}}$ for regular sensing.}
\label{fig:lena}
\end{figure*}

\subsection{Experiments with Lena Image}
\noindent\textbf{Experiment set-up}.
We evaluate the benefits of using 
$\bA^{(\textup{final})}_{\textup{preferential}}$ 
for the Lena image with dimension $512\times 512$. 
Notice that
the sensing matrix would have been prohibitively large
if we used the whole image
as the input. To put more
specifically, we would need a 
matrix with the width
$512^2 = 262144$. 
To handle such issue, we divide the whole images into a set of sub-blocks with 
dimensions $32\times 32$ and design one sensing matrix with 
the width $32^2 = 1024$. 
For each sub-block, we first obtain a sparse representation with a 2D Haar transform 
and then reconstruct the signal in \eqref{eq:haar_sense_relation}. 

\noindent\textbf{Discussion}. 
The comparison of results is plotted in Fig.~\ref{fig:lena}, from 
which we conclude that 
the sensing matrix 
$\bA^{(\textup{final})}_{\textup{preferential}}$ has much 
better performance in image reconstruction in comparison with the sensing matrix $\bA_{\textup{regular}}$.
The ratios $r_{H, (p)}$ and $r_{H, (r)}$ are computed as $0.0446
$ and $0.3029$, respectively; 
while the ratio $r_{W, (p)}$ and $r_{W, (r)}$ 
are computed as $0.0709$
and $0.3144$, respectively.  

\begin{remark}
The degree distributions
$\lambda_H(\cdot)$
and $\lambda_L(\cdot)$ of the 
variable nodes for the sensing matrix
$\bA^{(\textup{final})}_{\textup{preferential}}$ are obtained as 
\[
&\lambda_H(\alpha) \\
& = 0.0057856\alpha +0.025915\alpha^{2} +0.36394\alpha^{3} +0.35183\alpha^{4} 
\\ &+0.10333\alpha^{5} + 0.04134\alpha^{6} +0.021619\alpha^{7} +0.013508\alpha^{8} \\
& + 0.0094374\alpha^{9} +0.0070906\alpha^{10} 
+ 0.0056\alpha^{11}  \\
& +0.0045851\alpha^{12} + 0.0038574\alpha^{13} +0.0033145\alpha^{14} \\
&+0.0028963\alpha^{15}+
.0025659\alpha^{16} +0.0022992\alpha^{17} \\
& +0.0020801\alpha^{18} +0.0018973\alpha^{19} +0.0017428\alpha^{20} \\
&+ 0.0016109\alpha^{21} +0.001497\alpha^{22} +0.001398\alpha^{23} \\
& +0.0013111\alpha^{24} + 0.0012344\alpha^{25} 
+ 0.0011662\alpha^{26} \\
& +0.0011053\alpha^{27} +0.0010506\alpha^{28} + 0.0010013\alpha^{29} \\
& +0.0009565\alpha^{30} 
+ 0.00091576\alpha^{31} +0.00087852\alpha^{32} \\
&+ 0.00084436\alpha^{33} +0.00081292\alpha^{34} +0.00078388\alpha^{35} \\
&+ 0.00075697\alpha^{36} + 0.00073197\alpha^{37} +0.00070867\alpha^{38} \\
&+0.00068691\alpha^{39} +0.00066652\alpha^{40} + 0.00064738\alpha^{41} \\
& +0.00062937\alpha^{42} +0.00061238\alpha^{43} +0.00059633\alpha^{44} \\
&+ 0.00058114\alpha^{45} 
+ 0.00056673\alpha^{46} +0.00055304\alpha^{47} \\
& +0.00054001\alpha^{48} +0.0005276\alpha^{49}; \\
&\lambda_L(\alpha) =\alpha. 
\]
The check node degrees
$\deg{c}_{\mathsf{H}}$ 
and $\deg{c}_{\mathsf{L}}$ are both set as $4$. 
Meanwhile, the sensing matrix $\bA_{\textup{regular}}$ designed in 
\eqref{eqn:density_evolve_optim} is a regular sensing matrix whose 
variable node and check node degree distributions 
are given by 
$\lambda(\alpha) = \alpha^2$ and $\rho(\alpha) = \alpha^7$,
respectively.
\end{remark}

\section{Conclusions}
\label{sec:conclude}
This paper presented a general framework of the
sensing matrix design for a linear measurement 
system. 
Focusing on a sparse
sensing matrix $\bA$, we associated it with 
a graphical model $\calG = (\calV, \calE)$ and 
transformed the design of $\bA$ to 
the connectivity problem in $\calG$. 
With the density evolution technique, 
we proposed 
two design strategies, i.e.,
regular sensing and preferential sensing. 
In the regular sensing scenario, all entries of 
the signal are recovered with equal accuracy; while 
in the preferential sensing scenario, 
the entries in the high-priority sub-block 
are recovered more accurately (or exactly) 
relative to the entries in the low-priority sub-block.
We then analyzed the 
impact of the connectivity of 
the graph on the recovery performance. 
For the regular sensing, our framework 
can reproduce the classical result of Lasso, 
i.e., the number of measurements $m$ should be at least
in the order $O(k\log n)$, where 
$n$ is the 
length of the signal and
$k$ is the sparsity number. 
For the preferential sensing, our framework can lead to 
a significant reduction of the reconstruction error in the 
high-priority part and a modest reduction of the error in
the whole signal. 
Numerical experiments with both 
synthetic data and real-world data are 
presented to corroborate our claims. 


\begin{appendices}

\section{Proof of Thm.~\ref{thm:equal_lasso}}
\label{thm_proof:equal_lasso}
\begin{proof}
We begin the proof by restating the DE equation w.r.t. 
$E^{(t+1)}$ and $V^{(t+1)}$ as 

\vspace{-5mm}
{\footnotesize
\[
& E^{(t+1)} = \underbrace{\Expc_{\prior(s),z\sim \normdist(0, 1)}\Bracket{\prox\bracket{s + a_1z\sqrt{E^{(t)}} ; \beta a_2 V^{(t)}}- s}^2}_{\defequal \Psi_E\bracket{E^{(t)}; V^{(t)}}}; \notag \\
& V^{(t+1)} =\underbrace{\Expc_{\prior(s), z\sim \normdist(0, 1)}\Bracket{ \beta a_2 V^{(t)} \prox^{'}\bracket{s + a_1z\sqrt{E^{(t)}}; \beta a_2 V^{(t)} }}}_{\defequal \Psi_V\bracket{E^{(t)}; V^{(t)} }}.
\]
}

\par \vspace{-2mm}\noindent
The derivation of the necessary conditions for
$\lim_{t\rightarrow \infty} \bracket{E^{(t)}, V^{(t)}} = 
\bracket{0, 0}$ consists of two 
parts: 
\begin{itemize}
\item 
\textbf{Part I}.	
We verify that $(0, 0)$ is a fixed-point 
of the DE equation; 
\item 
\textbf{Part  II}. We consider the necessary condition such that DE equation converges 
within the proximity of the origin points, i.e., 
$E^{(t)}$ and $V^{(t)}$ are close to zero.
\end{itemize}
Since Part I can be easily verified, we put
our major focus on Part II. 
Define the difference across iterations 
as $\delta^{(t)}_{E} = E^{(t+1)} - E^{(t)}$ and 
$\delta^{(t)}_{V} = V^{(t+1)} - V^{(t)}$, we 
would like to show 
$\lim_{t\rightarrow \infty}(\delta_E^{(t)}, \delta_V^{(t)}) = \bracket{0, 0}$. 
With Taylor expansion, we obtain 

{\small \vspace{-2mm}
\begin{align}
\label{eq:equal_sparse_diff_update}
\delta^{(t+1)}_{E} =~& \Psi_E\bracket{ E^{(t+1)}, V^{(t+1)}} - \Psi_E\bracket{E^{(t)}, V^{(t)}} \notag \\
=~&\bracket{\frac{\partial \Psi_E\bracket{E, V}}{\partial E}\left|_{E = E^{(t)}, V = V^{(t)}} \right.}\cdot \delta_E^{(t)} \notag \\
+~&\bracket{\frac{\partial \Psi_E\bracket{E, V}}{\partial V}\left|_{E = E^{(t)}, V = V^{(t)}} \right.} \cdot \delta_V^{(t)}\notag \\
+~& O\bracket{\bracket{\delta_E^{(t)}}^2} + O\bracket{\bracket{\delta_V^{(t)}}^2}. 
\end{align}\vspace{-2mm}
}

\par \noindent
Consider the region where $\delta_E^{(t)}$
and $\delta_V^{(t)}$ are sufficiently 
small, we require 
$\delta_E^{(t)}$ and $\delta_V^{(t)}$ to converge to zero.  
Notice the quadratic terms in \eqref{eq:equal_sparse_diff_update}
can be safely omitted in this region.  
Denote the gradients $\bracket{\frac{\partial \Psi_E\bracket{E, V}}{\partial E}}^{(t)}\left|_{E = E^{(t)}, V = V^{(t)}} \right.$, 	$\frac{\partial \Psi_E\bracket{E, V}}{\partial V}\left|_{E = E^{(t)}, V = V^{(t)}} \right.$, 
	$\frac{\partial \Psi_V\bracket{E, V}}{\partial E}\left|_{E = E^{(t)}, V = V^{(t)}} \right.$, and
$\frac{\partial \Psi_V\bracket{E, V}}{\partial V}\left|_{E = E^{(t)}, V = V^{(t)}} \right.$
as $\bracket{\frac{\partial \Psi_E\bracket{E, V}}{\partial E}}^{(t)}$, $\bracket{ 	\frac{\partial \Psi_E\bracket{E, V}}{\partial V}}^{(t)}$, 
$\bracket{	\frac{\partial \Psi_V\bracket{E, V}}{\partial E}}^{(t)}$, and
$\bracket{\frac{\partial \Psi_V\bracket{E, V}}{\partial V}}^{(t)}
$, respectively.
We obtain the linear equation 
\[
\begin{bmatrix}
\delta_E^{(t+1)} \\
\delta_V^{(t+1)}
\end{bmatrix} =~& \
\underbrace{\begin{bmatrix}
	\bracket{\frac{\partial \Psi_E\bracket{E, V}}{\partial E}}^{(t)}& 
	\bracket{\frac{\partial \Psi_E\bracket{E, V}}{\partial V}}^{(t)} \\
	\bracket{\frac{\partial \Psi_V\bracket{E, V}}{\partial E}}^{(t)} & 
	\bracket{\frac{\partial \Psi_V\bracket{E, V}}{\partial V}}^{(t)}
\end{bmatrix}}_{\defequal \bL^{(t)}} \begin{bmatrix}
\delta_E^{(t)} \\
\delta_V^{(t)}
\end{bmatrix}, 
\]
and would require the lower bound of the operator norm 
of the matrix $\bL^{(t)}$ to be no greater than $1$, 
i.e., $\inf_t\opnorm{\bL^{(t)}}\leq 1$, since otherwise 
the values of $\delta_E^{(t)}$ and 
$\delta_V^{(t)}$ will keep increasing. 
Exploiting the fact $\frac{\partial \Psi_V\bracket{E, V}}{\partial E} = 0$, 
we conclude 
\[
\opnorm{\bL^{(t)}} = 
\max\Bracket{\bracket{\frac{\partial \Psi_E\bracket{E, V}}{\partial E}}^{(t)}, \bracket{\frac{\partial \Psi_V\bracket{E, V}}{\partial V}}^{(t)}}.
\]
The proof is then concluded by computing the lower bounds of the
gradients $\frac{\partial \Psi_E\bracket{E, V}}{\partial E}$ and 
$\frac{\partial \Psi_V\bracket{E, V}}{\partial V}$ as 

{\small \vspace{-2mm}
\begin{align}
\label{eq:equal_psie_lower_bound}
& \frac{\partial \Psi_E\bracket{E, V}}{\partial E}\left|_{E = E^{(t)}, V = V^{(t)}} \right. \notag \\
=~& 
a_1^2 \cdot \Expc_{\prior(s)}
\Bracket{\Phi\bracket{-\frac{s + a_2 V^{(t)}}{a_1\sqrt{E^{(t)}}}} + \Phi\bracket{\frac{s-a_2 V^{(t)}}{a_1\sqrt{E^{(t)}}}}} \notag  \\
\stackrel{\cirone}{=}~& \frac{a_1^2k}{n}\Bracket{\Phi\bracket{-\frac{c_0 + a_2 V^{(t)}}{a_2\sqrt{E^{(t)}}}} + \Phi\bracket{\frac{c_0 - a_2 V^{(t)}}{a_1\sqrt{E^{(t)}}}}} \notag \\
+~& 2a_1^2\bracket{1-\frac{k}{n}}\Phi\bracket{-\frac{a_2V^{(t)}}{a_1E^{(t)}}} \notag \\
\stackrel{\cirtwo}{\rightarrow}~& \frac{k a_1^2}{n} + 2a_1^2\bracket{1-\frac{k}{n}}
\Phi\bracket{- \frac{a_2 V^{(t)}}{\sqrt{a_1 E^{(t)}}}} \stackrel{\cirthree}{\geq} 
\frac{ka_1^2 }{n}; \notag \\
&\frac{\partial \Psi_V\bracket{E, V}}{\partial V}\left|_{E = E^{(t)}, V = V^{(t)}} \right. \notag \\
=~& 
\beta a_2 \cdot \Expc_{\prior(s)}
\Bracket{\Phi\bracket{-\frac{s + a_2 V^{(t)}}{a_1\sqrt{E^{(t)}}}} + \Phi\bracket{\frac{s-a_2 V^{(t)}}{a_1\sqrt{E^{(t)}}}}} \notag \\
\stackrel{\cirfour}{=}~&  \frac{\beta a_2 k}{n}\Bracket{\Phi\bracket{-\frac{c_0 + a_2 V^{(t)}}{a_1\sqrt{E^{(t)}}}} + \Phi\bracket{\frac{c_0 - a_2 V^{(t)}}{a_1\sqrt{E^{(t)}}}}} \notag \\
+~& 2\beta a_2\bracket{1-\frac{k}{n}}\Phi\bracket{-\frac{a_2V^{(t)}}{a_1E^{(t)}}} \notag  \\
\stackrel{\cirfive}{\rightarrow}~& 
\frac{k\beta a_2}{n} + 
2\beta a_2\bracket{1- \frac{k}{n}}\Phi\bracket{- \frac{a_2 V^{(t)}}{\sqrt{a_1 E^{(t)}}}}
\stackrel{\cirsix}{\geq} \frac{k\beta a_2}{n}, 
\end{align}
\vspace{-2mm} 
}

\par \noindent
where $\Phi(\cdot) = (2\pi)^{-1/2}\int_{-\infty}^{(\cdot)}e^{-z^2/2}dz$
is the CDF of the standard normal RV $z$, namely, 
$z\sim \normdist(0, 1)$. 
In $\cirone$ and $\cirfour$ we use the prior distribution 
$\prior(s) = k/n\cdot\Ind(c_0) + (1-k/n)\Ind(0)$. Further,
in $\cirtwo$ and $\cirfive$ we use the fact
\[
\lim_{E^{(t)}\rightarrow 0} 
\Phi\bracket{-\frac{c_0 + a_2 V^{(t)}}{\sqrt{a_1 E^{(t)}}}} 
+ \Phi\bracket{\frac{c_0 - a_2 V^{(t)}}{\sqrt{a_1 E^{(t)}}}} = 1,
\] 
since $c_0 \neq 0$. Finally, in 
$\cirthree$ and $\cirsix$ we omit the non-negative terms $\Phi(\cdot)$.

\end{proof}

\vspace{-10mm}
\iftrue 
\section{Example of regular sensing with a Gaussian prior}
\label{sec:gaussian_prior}
In addition to the Laplacian prior studied in 
Subsec.~\ref{subsec:equal_protect_laplacian}, 
we also investigate the Gaussian prior. 
Assuming the ground-truth $\bx^{\natural}$ to be 
Gaussian distributed with zero mean and unit variance,
we would like to recover the signal $\bx^{\natural}$
with the regularizer $f(\bx) = \norm{\bx}{2}^2$. 
In this case, the DE equation reduces to 
\begin{align}
\label{eq:ridge_de}
E^{(t+1)} &= \frac{a_1^2 E^{(t)} + a_2^2 (V^{(t)})^2}{\bracket{1 + a_2 V^{(t)}}^2}; \notag \\
V^{(t+1)} &= \frac{a_2 V^{(t)}}{1 + a_2 V^{(t)}}, 
\end{align}
where $a_1, a_2$ are defined the same as above. 
Then we have the following theorem.

\begin{figure}[!h]
\centering
\mbox{
\includegraphics[width = 1.65in]{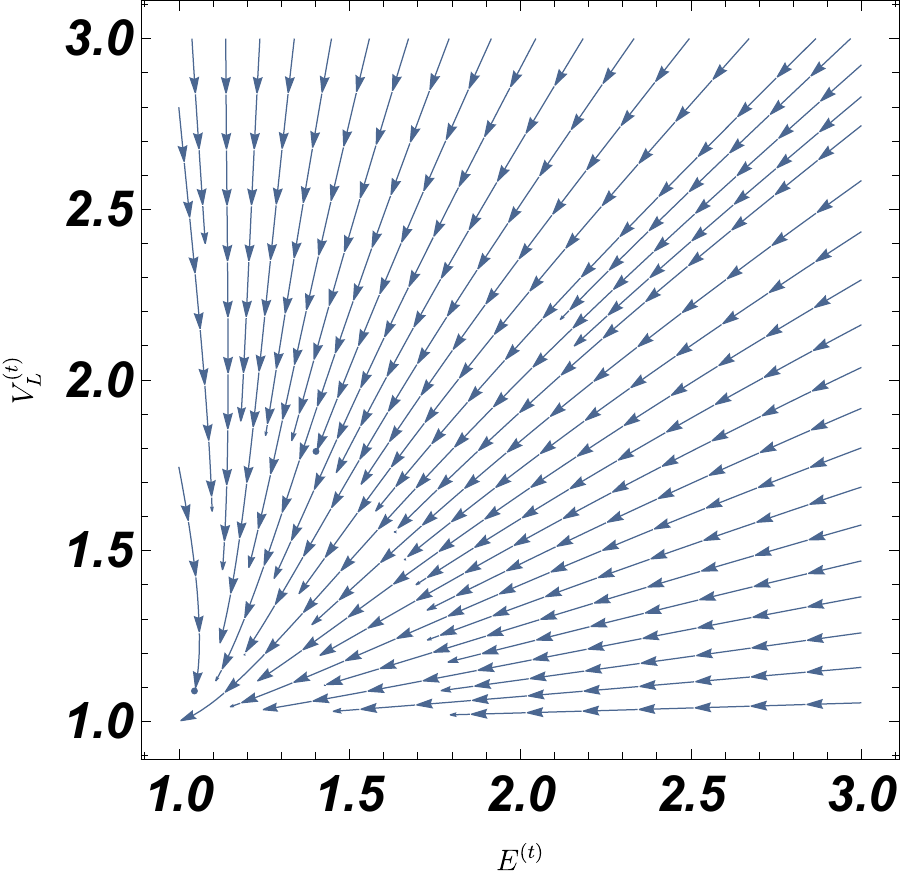}
\includegraphics[width = 1.65in]{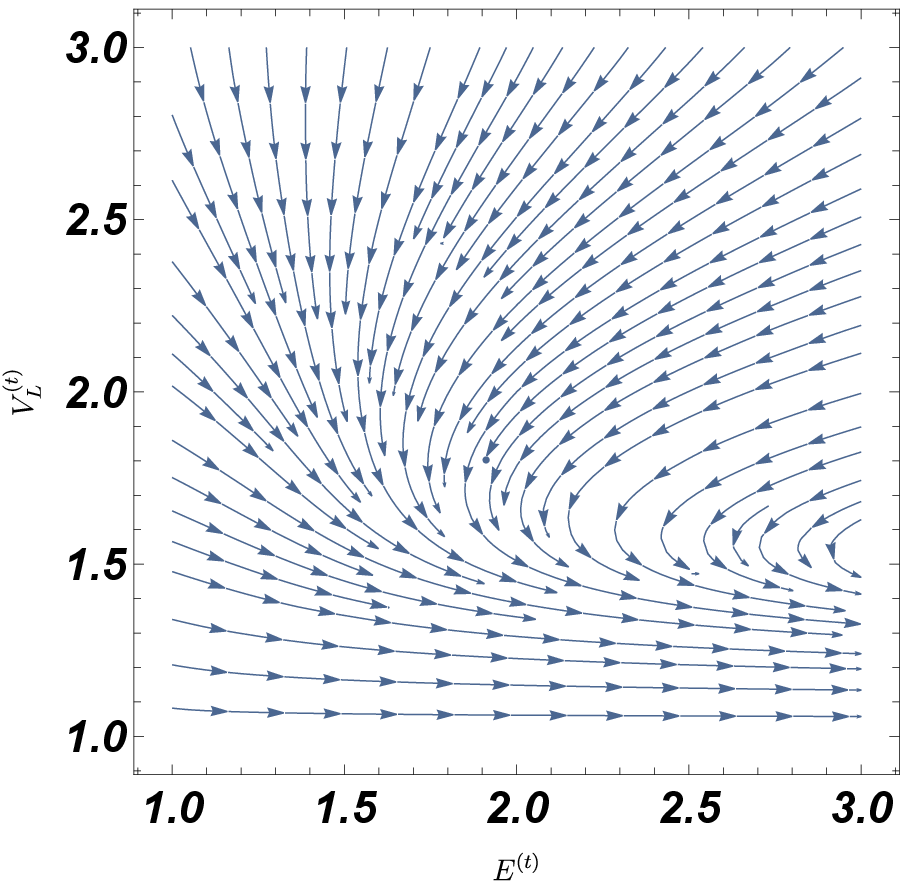}
}
\caption{Illustration of DE in 
\eqref{eq:ridge_de} when 
setting $\prior(s) = \Ind(s=1)$. 
\textbf{Left panel}: $\sum_{i,j}\rho_i \lambda_j\sqrt{i/j} < 1$. \textbf{Right panel}: 
$\sum_{i,j}\rho_i \lambda_j\sqrt{i/j} > 1$.
Notice that the left panel has a fix-point 
$(0, 0)$ while the right panel is with non-zero
fix-point. 
}
\label{fig:de_ridge}	
\end{figure}

\begin{theorem}
\label{thm:equal_ridge_regress}
Provided that $\sum_{i, j}\rho_i \lambda_j \sqrt{i/j} < 1$, the 
average error $E^{(t)}$ and the variance 
$V^{(t)}$ decrease exponentially 
after some iteration index $T$, that is, 
$E^{(t)} \leq e^{-c_0 (t-T)}E^{(T)}$ and
$V^{(t)} \leq e^{-c_1 (t-T)}V^{(T)}$
whenever $t\geq T$, where $c_0, c_1 > 0$ are some 
fixed constants.
\end{theorem}
\par \noindent
An illustration of the DE 
in \eqref{eq:ridge_de} is shown in Fig.~\ref{fig:de_ridge}.

\begin{proof}
We begin the proof by restating that the functions $\Expc_z h_{\mean}(\cdot; \cdot)$ 
and $\Expc_z h_{\var}\bracket{\mean; \var}$ are 
written as 
\[
\Expc_z h_{\mean}\bracket{s + a_1 z\sqrt{E^{(t)}}; a_2 V^{(t)}} &= \
\frac{a_1^2 E^{(t)}+ a_2^2 \bracket{V^{(t)}}^2 s^2}{\bracket{1 + a_2 V^{(t)}}^2}; \\
\Expc_z h_{\var}\bracket{s + a_1 z\sqrt{E^{(t)}}; a_2 V^{(t)}} &= \
\frac{a_2 V^{(t)}}{1 + a_2 V^{(t)}},
\]
which can be easily verified. 
Then we prove that $V^{(t)}$ decreases exponentially since  
$a_2 > 0$ and hence for an arbitrary time index $T_1$ the relation 
\[
V^{(t)} \leq \bracket{\frac{a_2}{1+a_2}}^{t-T_1} V^{(T_1)} = 
e^{-c_1\bracket{t-T_1}}V^{(T_1)} 
\]
holds 
for $t \geq T_1$, where $c_1$ is defined as $\log(1+a_2^{-1}) > 0$.

Afterwards, we study the behavior of $E^{(t)}$. Denote $V_S$ as 
$\Expc_{\prior(s)}(s^2)$, we have 
\begin{align}
\label{eq:de_ridge_E_update}
E^{(t+1)} \leq~& a_1^2 E^{(t)} + 
\frac{a_2 V_S V^{(t)}}{2} \notag \\
\stackrel{\cirone}{\leq}~&
a_1^2 E^{(t)} + \frac{a_2 V_S}{2}\bracket{\frac{a_2}{1+a_2}}^t V^{(0)}, 
\end{align}
where in $\cirone$ we use the relation 
$V^{(t)} \leq \bracket{a_2 /(1+a_2)}^t V^{(0)}$.
Define a new sequence 
$\wt{E}^{(t)} = E^{(t)}/a_1^{2t}$, 
we can transform \eqref{eq:de_ridge_E_update} 
to 
\[
\wt{E}^{(t+1)} =~&
\frac{E^{(t+1)}}{a_1^{2(t+1)}} \leq
\frac{E^{(t)}}{a_1^{2t}} + 
\frac{a_2 V_S V^{(0)}}{2a_1^2}
\bracket{\frac{a_2}{(1+a_2) a_1^2}}^t \\
=~& \wt{E}^{(t)} + \frac{a_2 V_S V^{(0)}}{2a_1^2}
\bracket{\frac{a_2}{(1+a_2) a_1^2}}^t, 
\]
after rearranging the terms. 
Due to the 
time-invariance, we also have the relation 
\[
\wt{E}^{(t)} \leq 
\wt{E}^{(t-1)} + \frac{a_2 V_S V^{(0)}}{2a_1^2}
\bracket{\frac{a_2}{(1+a_2) a_1^2}}^{t-1}. 
\]
Iterating over all such inequalities, we obtain 
the equation 

{\small \vspace{-2mm}
\begin{align*}
\wt{E}^{(t+1)}  \leq 
\wt{E}^{(1)} + 
\frac{a_2 V_S V^{(0)}}{2a_1^2}
\frac{\frac{a_2}{(1+a_2)a_1^2}\bracket{1 - \bracket{\frac{a_2}{(1+a_2)a_1^2}}^t}}{1 - \bracket{\frac{a_2}{(1+a_2)a_1^2}}}, 
\end{align*} \vspace{-2mm}
}

\par \noindent
which leads to 

{\small \vspace{-2mm}
\begin{align}
\label{eq:de_ridge_E_adv_update}
E^{(t+1)} \leq 
a_1^{2t} E^{(1)} + 
\underbrace{\frac{a_2 V_S V^{(0)}}{2a_1^2}\cdot 
\frac{a_2}{1+a_2}
\frac{a_1^{2t} - \bracket{\frac{a_2}{1+a_2}}^t}{1-\frac{a_2}{(1+a_2)a_1^2}}}_{I}. 
\end{align}
\vspace{-2mm}
}

\par \noindent
Since $a_1 < 1$ and $a_2/(1+a_2) < 1$, we have the second term 
$I$ in \eqref{eq:de_ridge_E_adv_update} to be negligible as $t$
goes to infinity. Hence we can choose a sufficiently large $T$ 
such that for $t\geq T$, we have $E^{(t+1)}$ is approximately 
equal to $a_1^{2t} E^{(1)}$ and conclude the exponential decay of 
$E^{(t)}$.  
\end{proof}

\fi 

\end{appendices}

\section*{Acknowledgment}
This material is based upon work supported by the National Science Foundation under Grant No. CCF-2007807 and ECCS-2027195.

\bibliographystyle{authordate3}
\bibliography{density_evolve, permute_var}


\newpage 
 \section{Discussion of the DE for both regular and 
irregular designs}
First we explain the physical meaning of the 
quantities $E^{(t)}$ and $V^{(t)}$, which 
track the average error and the average variance 
at the $t$th iteration, respectively. 
Since the physical meaning of $V^{(t)}$ can be easily obtained, 
we focus on the explanation of $E^{(t)}$.
For the convenience of the analysis, 
we rewrite the MAP estimator as

{\small \vspace{-4mm}
\[
\wh{\bx} = \argmax_{\bx} 
\exp\bracket{-\frac{\gamma \norm{\by - \bA\bx}{2}^2}{2\sigma^2 }}
\cdot \exp\bracket{-\gamma f(\bx)}, 
\]  \vspace{-2mm}
}

\par \noindent
where $\gamma > 0$ is a redundant positive constant. 
Then we restate the message-passing algorithm, which is used to solve the 
MAP estimator, as 
\[
\cvinfoidx{a}{i}{t+1}(x_i) &\cong \hspace{-1mm}\int\hspace{-2mm} \prod_{j\in \partial a\setminus i}  \vcinfoidx{j}{a}{t}(x_i) \times e^{-\frac{\gamma\bracket{y_a - \sum_{j=1}^n A_{aj}x_j}^2}{2\sigma^2}}dx_j \\
\vcinfoidx{i}{a}{t+1}(x_i) &\cong e^{-\gamma f(x_i)} \prod_{b\in \partial i \setminus a}
\cvinfoidx{b}{i}{t+1}(x_i).
\]
The MAP estimator of $\hat{x}_i$ is hence written as 
\[
\hat{x}_i = \argmax_{x_i} \Prob\bracket{x_i|\by} \approx 
\argmax_{x_i} e^{-\gamma f(x_i)}\prod_{a\in \partial i}
\cvinfoidx{a}{i}{t}(x_i).
\] 
Notice that  
$\hat{x}_i$ can be rewritten as the mean w.r.t. the probability measure 
$e^{-\gamma f(x_i)}\prod_{a\in \partial i}\cvinfoidx{a}{i}{t}$, namely,
\[
\hat{x}_i \approx \int_{x_i} x_i e^{-\gamma f(x_i)} \
\prod_{a\in \partial i}\cvinfoidx{a}{i}{t}(x_i) dx_i, 
\]
by letting $\gamma \rightarrow \infty$. 
Since the mean $\vcmean{i}{a}$ is computed as 
\[
\vcmean{i}{a} = 
\int_{x_i} x_i e^{-\gamma f(x_i)} \
\prod_{b\in \partial i\setminus a}\cvinfoidx{b}{i}{t}(x_i) dx_i, 
\]
which is close to $\hat{x}_i$, we obtain the 
approximation  
$m^{-1}\sum_{a=1}^m(\vcmean{i}{a} - x_i^{\natural})^2$
as $(\hat{x}_i  - x_i^{\natural})^2$. We then conclude 
\[
E^{(t)} = \frac{1}{mn}\sum_{i=1}^n \sum_{a=1}^m 
\bracket{\vcmean{i}{a} - x_i^{\natural}}^2 \approx 
\frac{1}{n}\sum_{i=1}^n \bracket{\hat{x}_i  - x_i^{\natural}}^2,
\]
which is approximately 
the average of error at the 
$t$th iteration. 
Having discussed the physical meaning of 
the quantities $E^{(t)}$ and $V^{(t)}$, we turn to the derivation of 
the DE equation. 

\vspace{-2mm}
\subsection{Supporting Lemmas}
\noindent We begin the derivation with the following lemma, which is stated as 
\begin{lemma}
\label{lemma:cv_message_flow_equation}
Consider   
the message flow $\cvinfoidx{a}{i}{t+1}$ from the check node $a$ to the 
variable node $i$ and approximate it as a Gaussian RV with 
mean $\cvmeanidx{a}{i}{t+1}$ and variance 
$\cvvaridx{a}{i}{t+1}$, i.e., 
$\cvinfoidx{a}{i}{t+1}\sim\normdist\bracket{\cvmeanidx{a}{i}{t+1}, \cvvaridx{a}{i}{t+1}}$. 
Then, we can obtain the following update equation
at the $(t+1)$th iteration

{\small \vspace{-2mm}
\[
\cvmeanidx{a}{i}{t+1} &=  x_i + A \sum_{j\in \partial a \setminus i}A_{ai}A_{aj}\bracket{x_j - \
\vcmeanidx{j}{a}{t} } + AA_{ai}w_{a}; \\
\cvvaridx{a}{i}{t+1} &= A\sigma^2 + |\partial a|V^{(t)}, 
\]\vspace{-2mm}
}

\par \noindent
where  $|\partial a|$ denotes
the degree of the check node $a$.
\end{lemma}

\vspace{-6mm}
\begin{proof}
Consider the message flow $\cvinfoidx{a}{i}{t+1}$ from check-node to variable node
at the $(t+1)$th iteration 

{\small \vspace{-2mm}
\begin{align}
\label{eq:de_derive_check_variable}
\cvinfoidx{a}{i}{t+1} =~&
\frac{1}{Z_{a\rightarrow i}^{t}} 
\int \prod_{j\in \partial a \setminus i} 
\vcinfoidx{j}{a}{t}(x_j) \notag \\
\times~& 
\exp\bracket{-\frac{\gamma \bracket{y_a - \sum_{j=1}A_{aj}x_j}^2
}{2\sigma^2}}dx_j. 
\end{align} \vspace{-2mm}
}

\par \noindent
Approximate the message flow 
$\vcinfoidx{j}{a}{t+1}$ as a Gaussian RV with 
mean $\vcmeanidx{j}{a}{t+1}$ and variance $\vcvaridx{j}{a}{t+1}$. 
Plugging into \eqref{eq:de_derive_check_variable} yields 

{\small \vspace{-2mm}
\begin{align}
\label{eq:de_derive_check_variable_step1}
\cvinfoidx{a}{i}{t+1} =~&
\frac{1}{Z_{a\rightarrow i}^{t}} 
\int \prod_{j\in \partial a \setminus i} 
\exp\Bracket{-\frac{\gamma\bracket{x_j - \vcmeanidx{j}{a}{t}}^2 }{2\vcvaridx{j}{a}{t+1}}} \notag \\
\times~& 
\exp\bracket{-\frac{\gamma \bracket{y_a - \sum_{j=1}A_{aj}x_j}^2
}{2\sigma^2}}dx_j. 
\end{align}\vspace{-2mm}
}

\par \noindent
The direct calculation of the above integral involves the
cross terms such as $A_{aj_1}A_{aj_2}x_{j_1}x_{j_2}$ ($j_1 \neq j_2$), which can be 
cumbersome. To handle this issue, 
we adopt the trick in \cite{nishimori2001statistical, krzakala2012probabilistic}, whose basic idea is to 
introduce a redundant 
variable $\omega$ and exploit the 
relation 
\[
e^{-\frac{t^2}{2\sigma^2}} = 
\frac{1}{\sqrt{2\pi \sigma^2}} \int e^{-\frac{\omega^2}{2\sigma^2} 
+ \frac{it\omega }{\sigma^2}} d\omega, 
\]  
where $t$ is an arbitrary number.
As such, we can transform \eqref{eq:de_derive_check_variable_step1} to 

{\small \vspace{-4mm}
\begin{align*}
\cvinfoidx{a}{i}{t+1}& \cong 
\int d\omega \prod_{j\in \partial a \setminus i} dx_j\cdot 
\exp\Bracket{-\frac{\gamma\bracket{x_j - \vcmeanidx{j}{a}{t}}^2 }{2\vcvaridx{j}{a}{t+1}}} \\
~& \times 
\exp\Bracket{-\frac{i\omega \gamma \bracket{y_a - \sum_{j=1}A_{aj}x_j}
}{\sigma^2}} \cdot 
\exp\Bracket{-\frac{\gamma \omega^2}{2\sigma^2}}, 
\end{align*} \vspace{-2mm}
}

\par \noindent
which diminishes the cross term $x_{j_1}x_{j_2}$ ($j_1\neq j_2$).
Rearranging the terms for each $x_j$, we can iteratively 
perform the integral such that

\vspace{-4mm}
{\small 
\begin{align*}
& \int dx_j \cdot 
\exp\bracket{-\frac{\gamma\bracket{x_j - \vcmeanidx{j}{a}{t}}^2 }{2\vcvaridx{j}{a}{t}}+\frac{i\omega \gamma A_{aj}x_j}
{\sigma^2}} \\
=~&
\sqrt{\frac{2\pi \vcvaridx{j}{a}{t}}{\gamma}}
\cdot 
\exp\Bracket{-\frac{\gamma  (\cvmeanidx{j}{a}{t})^2}{2\cvvaridx{j}{a}{t} }+\frac{
\vcvaridx{j}{a}{t}\left(\frac{\gamma
\vcmeanidx{j}{a}{t}}{ \vcvaridx{j}{a}{t}}+\frac{i \gamma  \omega  A_{aj}}{\sigma ^2}\right)^2}{2 \gamma }} \\
=~&
\sqrt{\frac{2\pi \vcvaridx{j}{a}{t}}{\gamma}}
\cdot\exp\bracket{-\frac{\gamma  \omega ^2 A_{aj}^2 \vcvaridx{j}{a}{t}}{2 \sigma ^4}+\frac{i \gamma  \omega  A_{aj} \vcmeanidx{j}{a}{t}}{\sigma ^2}}.
\end{align*}
}

\par \noindent 
With some algebraic manipulations, we can 
compute its mean $\cvmeanidx{a}{i}{t+1}$ and its 
variance $\cvvaridx{a}{i}{t+1}$ as 

{\small \vspace{-2mm}
\begin{align*}
\cvmeanidx{a}{i}{t+1} &= 
\frac{A_{ai}\bracket{y_a - \sum_{j\in \partial a \setminus i}
A_{aj}\vcmeanidx{j}{a}{t} }}{A_{ai}^2}; \\
\cvvaridx{a}{i}{t+1} &= \
\frac{\sigma^2 + \sum_{j\in \partial a \setminus i} 
A_{aj}^2 \vcvaridx{j}{a}{t}}{A_{ai}^2}.
\end{align*} \vspace{-2mm}
}

\par \noindent
The following analysis focuses on how to 
approximate these two values. 
We begin by the discussion w.r.t. the variance  $\cvvaridx{a}{i}{t+1}$. 
Note we have 
\[
\cvvaridx{a}{i}{t+1} 
\stackrel{\cirone}{\approx} A\sigma^2 +  \sum_{j\in \partial a \setminus i} 
\vcvaridx{j}{a}{t}, 
\]
where in $\cirone$ we use 
$A_{ai}^2 \approx \Expc\bracket{A_{ai}^2|A_{ai}\neq 0} = A^{-1}$
for $i\in \partial a$. 
As for the sum $\sum_{j\in \partial a\setminus i}\vcvaridx{j}{a}{t}$, 
we can view it to be randomly sampled from the set
of variances $\set{\vcvaridx{j}{a}{t}}$ and approximate
it as 
\[
\sum_{j\in \partial a\setminus i}\vcvaridx{j}{a}{t}
\approx \bracket{|\partial a| -1}V^{(t)} \approx 
|\partial a| V^{(t)}. 
\]
Notice that the variance is closely related with the 
check node degree $|\partial a|$. Having obtained
the variance $\cvvaridx{a}{i}{t+1}$, we turn to the 
mean $\cvmeanidx{a}{i}{t+1}$, which is computed as 

{\small \vspace{-2mm}
\begin{align*}
\cvmeanidx{a}{i}{t+1} =~& \frac{A_{ai}\bracket{y_a - \sum_{j\in \partial a \setminus i}
A_{aj}\vcmeanidx{j}{a}{t} }}{A_{ai}^2}\\
\stackrel{\cirtwo}{\approx}~& A A_{ai}
\bracket{A_{ai}x_i + \sum_{j\in \partial a \setminus i}A_{aj}\bracket{x_j - \
\vcmeanidx{j}{a}{t} } + w_{a}} \\
\stackrel{\cirthree}{\approx} ~& x_i + A \sum_{j\in \partial a \setminus i}A_{ai}A_{aj}\bracket{x_j - \
\vcmeanidx{j}{a}{t} } + AA_{ai}w_{a}, 
\end{align*}\vspace{-2mm}
}

\par \noindent
where in $\cirtwo$ and $\cirthree$ we use the approximation 
$A^{2}_{ai} \approx A^{-1}$ for $i\in \partial a$.

\end{proof}



\vspace{-10mm}
\subsection{Derivation of DE}

We study  the message flow 
$\vcinfoidx{i}{a}{t+1}$ from the variable node $i$ to the 
check node $a$
\[
\vcinfoidx{i}{a}{t+1} \cong e^{-\gamma f(x_i)}
\prod_{b\in \partial i\setminus a}
e^{-\frac{\gamma\bracket{x_i - \cvmeanidx{b}{i}{t+1}}^2}{2\cvvaridx{b}{i}{t+1}}}.
\]
To begin with, we study the product 
$\prod_{b\in \partial i\setminus a}
\exp\bigg(-\frac{\gamma\bracket{x_i - \cvmeanidx{b}{i}{t+1}}^2}{2\cvvaridx{b}{i}{t}}\bigg)$. Its variance 
$\tilvcvaridx{i}{a}{t+1}$ is approximately computed as 
\[
\frac{\gamma}{\tilvcvaridx{i}{a}{t+1}} \approx \
\sum_{b\in \partial i \setminus a}\frac{\gamma}{\cvvaridx{b}{i}{t+1}},
\]
which yields
\[
\tilvcvaridx{i}{a}{t+1} = \
\bracket{\frac{|\partial i| - 1}{
A\sigma^2 + |\partial a| V^{(t)}}
}^{-1} \approx 
\frac{A\sigma^2 + |\partial a|V^{(t)}}{|\partial i|}.
\]
Further, the mean $\tilvcmeanidx{i}{a}{t+1}$ is calculated as 

{\small \vspace{-2mm}
\begin{align*}
& \tilvcmeanidx{i}{a}{t+1} = \bracket{\sum_{b\in \partial i\setminus a}\frac{\cvmeanidx{b}{i}{t+1}}{\cvvaridx{b}{i}{t+1}}}/\
\bracket{\sum_{b\in \partial i \setminus a}\frac{1}{\cvvaridx{b}{i}{t+1}}}^{-1} \\
\stackrel{\cirone}{=}~& \
\frac{A\sigma^2 + |\partial a|V^{(t)}}{|\partial i|}\\
\times ~&
\bracket{\sum_{b\in \partial i \setminus a}
\frac{
x_i + A \sum_{j\in \partial b \setminus i}A_{bi}A_{bj}\bracket{x_j - \
\vcmeanidx{j}{b}{t} } + A A_{bi}w_{b}
}{A\sigma^2 + |\partial a|V^{(t)}} } \\
\approx ~& \
x_i + \frac{A}{|\partial i|}\
\Bracket{
\sum_{j\in \partial b \setminus i}A_{bi}A_{bj}\bracket{x_j - \
\vcmeanidx{j}{b}{t} } + \sum_{b\in \partial i \setminus a}A_{bi}w_{b}}, 
\end{align*} \vspace{-2mm}
}

\par \noindent
where in $\cirone$ we invoke Lemma~\ref{lemma:cv_message_flow_equation}. 
We then approximate the term 
$
\sum_{j\in \partial b \setminus i}A_{bi}A_{bj}\bracket{x_j - \
\vcmeanidx{j}{b}{t} }$ + $\sum_{b\in \partial i \setminus a}A_{bi}w_{b}$
as a Gaussian RV with its mean being calculated as 
\[
\Expc\Bracket{\sum_{b\in \partial i\setminus a}
\sum_{j\in \partial b \setminus i}A_{bi}A_{bj}\bracket{x_j - \
\vcmeanidx{j}{b}{t} } + \sum_{b\in \partial i \setminus a}A_{bi}w_{b}} = 0,
\]
and its variance as

{\small \vspace{-2mm}
\begin{align*}
& \Expc\Bracket{\sum_{b\in \partial i\setminus a}
\sum_{j\in \partial b \setminus i}A_{bi}A_{bj}\bracket{x_j - \
\vcmeanidx{j}{b}{t} } + \sum_{b\in \partial i \setminus a}A_{bi}w_{b}}^2\\
=~& \Expc\Bracket{\sum_{b\in \partial i\setminus a}
\sum_{j\in \partial b \setminus i}A_{bi}A_{bj}\bracket{x_j - \
\vcmeanidx{j}{b}{t} } }^2 
+
\Expc\Bracket{\sum_{b\in \partial i \setminus a}A_{bi}w_{b}}^2 \\
\approx~& A^{-2}|\partial i|\sum_{j\in \partial a \setminus i}
\bracket{x_j - \vcmeanidx{j}{b}{t}}^2+ 
A^{-1}\sigma^2 |\partial i| \\
\stackrel{\cirtwo}{\approx}~&
|\partial i|
\bracket{A^{-2}|\partial a| E^{(t)} + A^{-1}\sigma^2}.
\end{align*}
\vspace{-2mm}
}

\par \noindent
In $\cirtwo$ we assume the 
term $\bracket{x_j - \vcmeanidx{j}{b}{t}}^2$ is randomly 
sampled among all possible pairs $(i, a)$. 
Hence for the fixed degree $|\partial i|$ and 
$|\partial a|$, we can approximate the mean $\tilvcmeanidx{i}{a}{t+1}$
as a Gaussian RV with mean $x_i + z\sqrt{\bracket{A\sigma^2 + |\partial a|E^{(t)}}/|\partial i|}$ and variance 
$\bracket{A\sigma^2+ |\partial a|V^{(t)}}/|\partial i|$, namely, 
\[
x\sim \normdist\bracket{x_i + z\sqrt{\frac{A\sigma^2 + |\partial a|E^{(t)}}{|\partial i|}},\frac{A\sigma^2+ |\partial a|V^{(t)}}{|\partial i|}},
\] 
where $z$ is a standard normal RV. 
Recalling that the distribution of the degrees of the variable node 
$i$ and check node $a$ satisfies 
 $\Prob(|\partial i| = \alpha) = \lambda_{\alpha}$ 
 and $\Prob\bracket{|\partial a| = \beta} = \rho_{\beta}$,
 we can approximate the distribution of the 
 product $\prod_{b\in \partial i \setminus a}\exp\bigg[-\gamma\big(x_i - \cvmeanidx{b}{i}{t}\big)^2/(2 \cvvaridx{b}{i}{t})\bigg]$ as the 
 mixture Gaussian $\sum_{i,j}\rho_i \lambda_j 
 \normdist\bracket{
z\sqrt{\frac{iE^{(t)} + A\sigma^2}{j}}, 
\frac{A\sigma^2 + i V^{(t)}}{j}}$ \footnote{
One hidden assumption is that there is no-local 
loops in the graphical model we constructed, which is
widely used in the previous work \cite{mezard2009information}.
} and further
approximate it as a single Gaussian RV 
with mean $x_i + \sum_{i,j}\rho_i \lambda_j z\sqrt{\frac{iE^{(t)} + A\sigma^2}{j}}$ and 
variance $\sum_{i, j} \rho_i \lambda_j \frac{A\sigma^2 + i V^{(t)}}{j}$.  
Invoking the definitions of 
$h_{\mean}(\cdot; \cdot)$ and 
$h_{\var}\bracket{\cdot; \cdot}$ as in \eqref{eq:hmeanvar_def}, 
we then approximate the mean $\vcmeanidx{i}{a}{t+1}$ and
the variance $\vcvaridx{i}{a}{t+1}$ as 
\[
\vcmeanidx{i}{a}{t+1} \approx 
 h_{\mean}\bigg(x_i + &
z\sum_{i,j}\rho_i \lambda_j \sqrt{\frac{iE^{(t)} + A\sigma^2}{j}} ;\\
 & \sum_{i, j} \rho_i \lambda_j \frac{A\sigma^2 + i V^{(t)}}{j}\bigg); \\
\vcvaridx{i}{a}{t+1} \approx  
h_{\var}\bigg(x_i + &
z\sum_{i,j}\rho_i \lambda_j \sqrt{\frac{iE^{(t)} + A\sigma^2}{j}} ; \\ & \sum_{i, j} \rho_i \lambda_j \frac{A\sigma^2 + i V^{(t)}}{j}\bigg).
\]
Then, the DE w.r.t. the average error $E^{(t+1)}$ is derived as 
\[
& E^{(t+1)} = \frac{1}{mn}\sum_{a=1}^m \sum_{i=1}^n 
\bracket{\vcmeanidx{i}{a}{t+1} - x^{\natural}_i}^{2}\\ 
\approx~&\Expc_{\prior(s)}\Expc_z 
\bigg[h_{\mean}\bigg(x^{\natural}_i + 
z\sum_{i,j}\rho_i \lambda_j \sqrt{\frac{iE^{(t)} + A\sigma^2}{j}} ;\\ & \quad \quad \quad \quad \quad \quad \quad \quad \quad \sum_{i, j} \rho_i \lambda_j \frac{A\sigma^2 + i V^{(t)}}{j}\bigg)-x_i^{\natural}\bigg]^2.
\] 
Following a similar method, we obtain the DE w.r.t. 
the average variance $V^{(t+1)}$ as stated in 
\eqref{eq:de_equal_protect}. This completes the proof. 

\subsection{Derivation of DE for Irregular Design}
Different from the regular design, we separately 
track the average error and average variance w.r.t. 
the high-priority part and low-priority part. 
Then we define four quantities, namely, $E_L^{(t)}, E_H^{(t)}$, 
$V_L^{(t)}$, and $V_H^{(t)}$, which are written as 

{\small \vspace{-2mm}
\begin{align*}
E_L^{(t)} &= \frac{1}{m n_L}\sum_{a=1}^m \sum_{i\in L}
\bracket{\vcmeanidx{i}{a}{t} - x_i^{\natural}}^2; \\
E_H^{(t)} &= \frac{1}{m n_H}\sum_{a=1}^m \sum_{i\in H}
\bracket{\vcmeanidx{i}{a}{t} - x_i^{\natural}}^2;  \\
V_L^{(t)} &= \frac{1}{m n_L}\sum_{a=1}^m \sum_{i\in L}\vcvaridx{i}{a}{t};\\ 
V_H^{(t)} &= \frac{1}{m n_H}\sum_{a=1}^m \sum_{i\in H}\vcvaridx{i}{a}{t},
\end{align*}\vspace{-2mm}
}

\par \noindent
where $n_H$ and $n_L$ denote the length of the 
high-priority part $\bx^{\natural}_H$ and low-priority 
part $\bx^{\natural}_L$, respectively. 
Following the same procedure as above then yields the proof 
of \eqref{eq:unequal_de}. The derivation details are omitted for the
 clarify of presentation.

\section{Discussion of Subsec.~\ref{subsec:preferential_laplace_prior}}
\label{sec:discuss_unequal}
We start the discussion by outlining the DE equation 
w.r.t. $E_H^{(t)}, E_L^{(t)}, V_H^{(t)},$ and $V_L^{(t)}$

{\small \vspace{-2mm}
\begin{align*}
E_H^{(t+1)} &= \Expc_{\prior(s)}\Expc_{z\sim\normdist(0, 1)}
\Bracket{\prox\bracket{s + z \cdot b_{H, 1}^{(t)}; \beta_H b_{H, 2}^{(t)} } - s}^2 \\
& \defequal   \Psi_{E, H}
\bracket{E_H^{(t)}, E_L^{(t)}, V_H^{(t)}, V_L^{(t)}};\notag  \\
E_L^{(t+1)} &= \Expc_{\prior(s)}\Expc_{z\sim\normdist(0, 1)}
\Bracket{\prox\bracket{s + z \cdot b_{L, 1}^{(t)}; 
\beta_L b_{L, 2}^{(t)} } - s}^2 \\
& \defequal \Psi_{E, L}\bracket{E_H^{(t)}, E_L^{(t)}, V_H^{(t)}, V_L^{(t)}}; \notag \\
V_H^{(t+1)} &= \Expc_{\prior(s)}\Expc_{z\sim \normdist(0,1)}\\
& \Bracket{\beta_H b_{H, 2}\cdot \prox^{'}\bracket{ s + z \cdot b_{H, 1}^{(t)}; 
\beta_H b_{H, 2}^{(t)} }} \\
& \defequal \Psi_{V, H}\bracket{E_H^{(t)}, E_L^{(t)}, V_H^{(t)}, V_L^{(t)}}; \notag \\
V_L^{(t+1)} &= \Expc_{\prior(s)}\Expc_{z\sim \normdist(0,1)}\\
& \Bracket{\beta_L b_{L, 2}\cdot \prox^{'}\bracket{ s + z \cdot b_{L, 1}^{(t)}; 
\beta_L b^{(t)}_{L, 2} } } \\
& \defequal \Psi_{V, L}\bracket{E_H^{(t)}, E_L^{(t)}, V_H^{(t)}, V_L^{(t)}}, 
\end{align*} \vspace{-4mm}
}

\par \noindent
where notation
$\prox(a; b)
$ is the soft-thresholding estimator defined as $\sign(a)\max(|a| - b, 0)$, notation
$\prox^{'}(a; b)$ is the derivative w.r.t. the first argument, 
and the notations
$b_{H, 1}^{(t)}, b_{H, 2}^{(t)}, b_{L,1}^{(t)}$, and $b_{L,2}^{(t)}$ are defined as 

{\small \vspace{-2mm}
\begin{align*}
b_{H, 1}^{(t)}& = \sum_{\ell, i, j} \lambda_{H,\ell} \rho_{H, i} \rho^L_j 
\sqrt{\frac{A\sigma^2 + iE_H^{(t)} + j E^{(t)}_L }{\ell}}; \\
b_{H, 2}^{(t)} &= \sum_{\ell, i, j} \lambda_{H,\ell} \rho_{H, i} \rho^L_j 
\frac{A\sigma^2 + i V^{(t)}_H + j V^{(t)}_L}{\ell}; \\
b_{L, 1}^{(t)} & = \sum_{\ell, i, j} \lambda^L_{\ell} \rho_{L, i} \rho_{H,j} 
\sqrt{\frac{A\sigma^2 + iE^{(t)}_L+ j E^{(t)}_H }{\ell}}; \\
b_{L, 2}^{(t)} &= \sum_{\ell, i, j} \lambda^L_{\ell} \rho_{L, i} \rho_{H,j} 
\frac{A\sigma^2 + iV^{(t)}_L+ jV^{(t)}_H }{\ell}.
\end{align*}
\vspace{-2mm}}

\par \noindent
Similar to the proof in Sec.~\ref{thm_proof:equal_lasso}, we define 
the differences across iterations as 
\[
\delta^{(t)}_{E, H} &\defequal E^{(t+1)}_H - E^{(t)}_H;~~\delta^{(t)}_{E, L} \defequal E^{(t+1)}_L - E^{(t)}_L; \\
\delta^{(t)}_{V, H} &\defequal V^{(t+1)}_H - V^{(t)}_H;~~ \delta^{(t)}_{V, L} \defequal V^{(t+1)}_H - V^{(t)}_H.
\]

\vspace{-5mm}
\subsection{Discussion of Eq.~\eqref{eq:unequal_variance}}
This subsection follows the same logic as in 
Sec.~\ref{thm_proof:equal_lasso}. 
We first relax the Requirement~\ref{require:noiseless_unequal} w.r.t. the average variance $V^{(t)}_H$ and $V^{(t)}_L$. Performing 
the Taylor-expansion, we obtain 

{\small \vspace{-2mm}
\begin{align} 
\label{eq:unequal_varaince_taylor_expansion}
& \delta_{V ,H}^{(t+1)}\notag  \\
=~&
\Psi_{V, H}\bracket{V_H^{(t+1)}, V_L^{(t+1)}, E_H^{(t+1)},
E_L^{(t+1)}} \notag \\
- ~& \Psi_{V, H}\bracket{V_H^{(t)}, V_L^{(t)}, E_H^{(t)},
E_L^{(t)}}\notag 
\\
{=}~& 
\bracket{\frac{\partial \Psi_{V, H}\bracket{\cdot}}{\partial E_H}\left|_{E_H = E_H^{(t)}, E_L = E_L^{(t)}, V_H = V_H^{(t)}, V_L = V_L^{(t)} } \right.}  
\delta_{E ,H}^{(t)} \notag \\
+~& 
\bracket{\frac{\partial \Psi_{V, H}\bracket{\cdot}}{\partial E_L} \left|_{E_H = E_H^{(t)}, E_L = E_L^{(t)}, V_H = V_H^{(t)}, V_L = V_L^{(t)} } \right.}
\delta_{E ,L}^{(t)}\notag \\
+~&\bracket{\frac{\partial \Psi_{V, H}\bracket{\cdot}}{\partial V_H}\left|_{E_H = E_H^{(t)}, E_L = E_L^{(t)}, V_H = V_H^{(t)}, V_L = V_L^{(t)} } \right.}  
\delta_{V ,H}^{(t)} \notag \\
+~& 
\bracket{\frac{\partial \Psi_{V, H}\bracket{\cdot}}{\partial E_H} \left|_{E_H = E_H^{(t)}, E_L = E_L^{(t)}, V_H = V_H^{(t)}, V_L = V_L^{(t)} } \right.}
\delta_{V ,L}^{(t)} \notag \\
+~& O\bracket{\bracket{\delta_{V ,H}^{(t)} }^2} + 
O\bracket{\bracket{\delta_{V ,L}^{(t)} }^2}.
\end{align}
\vspace{-2mm}
}

\par \noindent 
Following the same logic in Sec.~\ref{thm_proof:equal_lasso}, 
our derivation consists of two parts:
\begin{itemize}
\item 
\textbf{Part I}. We verify that $(0, 0)$ is a fixed 
point of the DE equation w.r.t. $V^{(t)}_H$
and $V^{(t)}_L$; 
\item 
\textbf{Part II}. We show the DE equation w.r.t. $V^{(t)}_H$ and $V^{(t)}_L$
converges within the proximity of the origin points. 
\end{itemize}

\vspace{-2mm} \par \noindent
Our following derivation focuses on showing 
that DE converges, or equivalently, 
$\lim_{t\rightarrow \infty}\bracket{\delta^{(t)}_{V, H}, \delta^{(t)}_{V, L} } = \bracket{0, 0}$, as 
the second part can be easily verified. 
We consider the region where $V^{(t)}_H, V^{(t)}_L, \delta^{(t)}_{V, H}$,
and $\delta^{(t)}_{V, L}$ are sufficiently small and hence can safely omit
the quadratic terms in \eqref{eq:unequal_varaince_taylor_expansion}. 
Exploiting the fact that $\partial \Psi_{V, H}/\partial E_H = 0$ and $ 
\partial \Psi_{V, H}/\partial E_L = 0$,
we obtain the linear relation 
\[
\begin{bmatrix}
\delta_{V ,H}^{(t+1)} \\
\delta_{V ,L}^{(t+1)}
\end{bmatrix} = 
\underbrace{\begin{bmatrix}
\bracket{\frac{\partial \Psi_{V, H}\bracket{\cdot}}{\partial V_H}}^{(t)} & \
\bracket{\frac{\partial \Psi_{V, H}\bracket{\cdot}}{\partial V_L}}^{(t)} 
 \\
 \bracket{\frac{\partial \Psi_{V, L}\bracket{\cdot}}{\partial V_H}}^{(t)}  & \
\bracket{\frac{\partial \Psi_{V, L}\bracket{\cdot}}{\partial V_L}}^{(t)} 	
\end{bmatrix}}_{\bL^{(t)}_{V}}
\begin{bmatrix}
\delta_{V ,H}^{(t)} \\
\delta_{V ,L}^{(t)}
\end{bmatrix}, 
\]
where the notation $\bracket{\frac{\partial \Psi_{V, H}\bracket{\cdot}}{\partial V_H}}^{(t)}$ is an abbreviation for the gradient 

\vspace{-4mm}
{\small 
\[
\bracket{\frac{\partial \Psi_{V, H}\bracket{\cdot}}{\partial V_H}}^{(t)} = 
\frac{\partial \Psi_{V, H}\bracket{\cdot}}{\partial V_H}\left|_{E_H = E_H^{(t)}, E_L = E_L^{(t)}, V_H = V_H^{(t)}, V_L = V_L^{(t)} } \right. .
\]
}

\par \vspace{-4mm} \noindent
Similarly we define the notations
$\bracket{{\partial \Psi_{V, H}\bracket{\cdot}}/{\partial V_L}}^{(t)}$, 
$\bracket{{\partial \Psi_{V, L}\bracket{\cdot}}/{\partial V_H}}^{(t)}$,
and
$\bracket{{\partial \Psi_{V, L}\bracket{\cdot}}/{\partial V_L}}^{(t)}$. 
Then we require $\inf_t \opnorm{\bL_V^{(t)}} \leq 1$. Otherwise, 
the values of $\delta^{(t)}_{V, H}$ and $\delta^{(t)}_{V, L}$
will keep increasing and stay away from zero. 
We then lower bound the gradients $
\bracket{\partial \Psi_{V, H}(\cdot)/\partial V_H}^{(t)}$ and
$\bracket{\partial \Psi_{V, H}(\cdot)/\partial V_L}^{(t)}$ as 

{\small \vspace{-2mm}
\begin{align*}
& \bracket{\frac{\partial \Psi_{V, H}(\cdot)}{\partial V_H}}^{(t)}
\stackrel{\cirone}{=} \beta_H 
\bracket{\sum_{\ell}\frac{\lambda_{H,\ell}}{\ell}}
\cdot \bracket{\sum_i i \rho_{H, i}} \\
\times ~&
\Bracket{2\bracket{1 - \frac{k_H}{n_H}}\Phi\bracket{-\frac{ \beta_H b^{(t)}_{H, 2}}{b^{(t)}_{H, 1}}}
+ \frac{k_H}{n_H}} \\
\stackrel{\cirtwo}{\geq}~& \frac{k_H \beta_H}{n_H}\bracket{\sum_{\ell}\frac{\lambda_{H,\ell}}{\ell}}
\cdot \bracket{\sum_i i \rho_{H, i}}; \\
&\bracket{\frac{\partial \Psi_{V, H}(\cdot)}{\partial V_L} }^{(t)}
\stackrel{\cirthree}{=}\beta_H\bracket{\sum_{\ell}\frac{\lambda_{H,\ell}}{\ell}}
\cdot \bracket{\sum_i i \rho_{L, i}} \\
\times~& 
\Bracket{2\bracket{1 - \frac{k_H}{n_H}}\Phi\bracket{-\frac{ \beta_H b^{(t)}_{H, 2}}{b^{(t)}_{H, 1}}}
+ \frac{k_H}{n_H}} \\
\stackrel{\cirfour}{\geq} ~& \frac{k_H\beta_H}{n_H}\bracket{\sum_{\ell}\frac{\lambda_{H,\ell}}{\ell}}
\cdot \bracket{\sum_i i \rho_{L, i}}, 
\end{align*}\vspace{-2mm}
} 

\par \noindent
where $\Phi(\cdot) = (2\pi)^{-1/2}\int_{-\infty}^{(\cdot)}e^{-z^2/2}dz$
is the CDF of the standard normal RV $z$, i.e., 
$z\sim \normdist(0, 1)$.   
In $\cirone$ and $\cirthree$, we follow the 
same computation procedure as in 
\eqref{eq:equal_psie_lower_bound}, and in 
$\cirtwo$ and $\cirfour$ we drop the non-negative terms $\Phi(\cdot)$. 
Following a similar procedure, we lower bound 
the gradients 
$\bracket{\partial \Psi_{V, L}(\cdot)/\partial V_H}^{(t)}$ and $
\bracket{\partial \Psi_{V, L}(\cdot)/\partial V_L}^{(t)}$ as 

{\small \vspace{-2mm}
\begin{align*}
\bracket{\frac{\partial \Psi_{V, L}(\cdot)}{\partial V_H}}^{(t)}
\geq ~& \frac{k_L\beta_L}{n_L}\bracket{\sum_{\ell}\frac{\lambda_{L,\ell}}{\ell}}
\cdot \bracket{\sum_i i \rho_{H, i}}; \\
\bracket{\frac{\partial \Psi_{V, L}(\cdot)}{\partial V_L} }^{(t)}
\geq  ~& \frac{k_L\beta_L}{n_L}\bracket{\sum_{\ell}\frac{\lambda_{L,\ell}}{\ell}}
\cdot \bracket{\sum_i i \rho_{L, i}}, 
\end{align*}\vspace{-2mm}
}

\par \noindent
and conclude the discussion.

\vspace{-4mm}
\subsection{Discussion of Eq.~\eqref{eq:unequal_error}}
This subsection relaxes the 
requirement $\lim_{t\rightarrow \infty}E^{(t)}_H = 0$, 
which consists of two parts: 
\begin{itemize}
	\item 
\textbf{Part I.}
we consider the necessary conditions such that DE equation w.r.t. $E^{(t)}_H$
converges; 
\item 
\textbf{Part II}. 
We verify that  
$0$ is a fixed point of DE w.r.t. $E^{(t)}_H$
given that $\lim_{t\rightarrow\infty}\bracket{V^{(t)}_H, V^{(t)}_L} = \bracket{0, 0}$. 
\end{itemize}
Since the second part can be easily verified, 
we focus on the first part. 
We consider the
region where 
 $E_H^{(t)}$ and $\delta^{(t)}_{E, H}$ are all 
 sufficiently small 
 and require $\delta_{E, H}^{(t)}$ 
to converge to zero. 
Via the Taylor expansion, we obtain the following linear equation 

\vspace{-2mm}
{\small 
 \begin{align}
 \label{eq:unequal_err_high_taylor_expand}
 \delta^{(t+1)}_{E, H} \approx  
 \bracket{\frac{\Psi_{E, H}(\cdot)}{\partial E_H}}^{(t)} \delta^{(t)}_{E, H} + \
 \bracket{\frac{\Psi_{E, H}(\cdot)}{\partial E_L}}^{(t)} \delta^{(t)}_{E, L}, 
 \end{align}
 }
 
\par \vspace{-4mm}  \noindent
where $ \bracket{\frac{\Psi_{E, H}(\cdot)}{\partial E_H}}^{(t)} $ denotes the 
 gradient $\frac{\Psi_{E, H}(\cdot)}{\partial E_H}$ 
 at the point $\bracket{E_H^{(t)}, E_L^{(t)}, V_H^{(t)}, V_L^{(t)}}$. 
Enforcing the variable $\delta^{(t)}_{E, H}$ to 
converge to zero, we require 
\[
\inf_t \Bracket{
 \bracket{\frac{\Psi_{E, H}(\cdot)}{\partial E_H}}^{(t)}
}^2 + 
\Bracket{\bracket{\frac{\Psi_{E, H}(\cdot)}{\partial E_L}}^{(t)}}^2 \leq 1. 
\]
Then our goal becomes lower-bounding 
the gradients, which are written as  

{\small \vspace{-2mm}
\begin{align}
\label{eq:unequal_error_relax_one}
& \bracket{\frac{\Psi_{E, H}(\cdot)}{\partial E_H}}^{(t)} 
\geq
\frac{k_H b^{(t)}_{H,1}}{n_H}
 \bracket{\sum_{\ell} \frac{\lambda_{H, \ell}}{\sqrt{\ell}}}
\bracket{\sum_{i, j}\frac{i\rho_{H, i} \rho_{L, j}}{\sqrt{iE^{(t)}_H + jE^{(t)}_L}} };
 \\
& \bracket{\frac{\Psi_{E, H}(\cdot)}{\partial E_L}}^{(t)}
\geq  
\frac{k_H b^{(t)}_{H,1}}{n_H}
 \bracket{\sum_{\ell} \frac{\lambda_{H, \ell}}{\sqrt{\ell}}}
\bracket{\sum_{i, j}\frac{j\rho_{H, i} \rho_{L, j}}{\sqrt{iE^{(t)}_H + jE^{(t)}_L}} }.
\label{eq:unequal_error_relax_two}
\end{align}\vspace{-2mm}
}

\par \noindent
Taking the limit $E_H^{(t)}\rightarrow 0$, we can 
conclude the relaxation by simplifying  
\eqref{eq:unequal_error_relax_one} and 
\eqref{eq:unequal_error_relax_two} as 

{\small \vspace{-2mm}
\begin{align*}
\bracket{\frac{\Psi_{E, H}(\cdot)}{\partial E_H}}^{(t)}  \geq \frac{k_H}{n_H}
 \bracket{\sum_{\ell} \frac{\lambda_{H, \ell}}{\sqrt{\ell}}}^2
\bracket{\sum_i \sqrt{i} \rho_{H, i}}^2; \\
\bracket{\frac{\Psi_{E, H}(\cdot)}{\partial E_L}}^{(t)} \geq 
\frac{k_H}{n_H}
 \bracket{\sum_{\ell} \frac{\lambda_{H, \ell}}{\sqrt{\ell}}}^2
\bracket{\sum_i \sqrt{i} \rho_{L, i}}^2.
\end{align*} \vspace{-2mm}
}

\vspace{-4mm}
\subsection{Discussion of Eq.~\eqref{eq:unequal_error_decrease_faster}}
The basic idea is to linearize the 
DE update equation with Taylor expansion 
and enforce the difference 
$\delta^{(t)}_{V, H}$ to decrease at a faster rate than 
$\delta^{(t)}_{V, L}$:

{\small \vspace{-2mm}
\begin{align}
\label{eq:unequal_noisy_decrease_faster}
\bracket{\frac{\Psi_{E, H}(\cdot)}{\partial E_H}}^{(t)}  
\leq 
\bracket{\frac{\Psi_{E, L}(\cdot)}{\partial E_H}}^{(t)}; \notag \\
\bracket{\frac{\Psi_{E, H}(\cdot)}{\partial E_L}}^{(t)}  
\leq 
\bracket{\frac{\Psi_{E, L}(\cdot)}{\partial E_L}}^{(t)}. 
\end{align}\vspace{-2mm}
}

\par \noindent
Following the same logic as \eqref{eq:unequal_error_relax_one} and 
\eqref{eq:unequal_error_relax_two}, we 
can lower-bound the gradients 
$\bracket{\frac{\Psi_{E, L}(\cdot)}{\partial E_H}}^{(t)}$ and $\bracket{\frac{\Psi_{E, L}(\cdot)}{\partial E_L}}^{(t)}$ 
as 

{\small \vspace{-2mm}
\begin{align*}
\bracket{\frac{\Psi_{E, L}(\cdot)}{\partial E_H}}^{(t)}  \geq \frac{k_L}{n_L}
 \bracket{\sum_{\ell} \frac{\lambda_{L, \ell}}{\sqrt{\ell}}}^2
\bracket{\sum_i \sqrt{i} \rho_{H, i}}^2; \\
\bracket{\frac{\Psi_{E, L}(\cdot)}{\partial E_L}}^{(t)} \geq 
\frac{k_L}{n_L}
 \bracket{\sum_{\ell} \frac{\lambda_{L, \ell}}{\sqrt{\ell}}}^2
\bracket{\sum_i \sqrt{i} \rho_{L, i}}^2.
\end{align*} \vspace{-2mm}
}

\par \noindent
Combining with \eqref{eq:unequal_noisy_decrease_faster} will then
yield the Requirement \ref{require:noisy_unequal}.

\end{document}